\documentclass[%
 reprint,
 superscriptaddress,
nofootinbib,
 amsmath,amssymb,
aip,
jcp,
]{revtex4-2}

\usepackage[utf8]{inputenc}
\usepackage{amsfonts}
\usepackage{amsmath,amssymb}
\usepackage{amsthm}
\usepackage{ascmac}
\usepackage{braket}
\usepackage{CJK}
\usepackage{comment}
\usepackage{here}
\usepackage{graphicx}% Include figure files
\usepackage{dcolumn}% Align table columns on decimal point
\usepackage{bm}% bold math
\usepackage{mathrsfs}
\usepackage{mathtools}
\usepackage[Symbol]{upgreek}
\usepackage{hyperref}% add hypertext capabilities
\hypersetup{% hyperref options
setpagesize=false,
bookmarksnumbered=false,%
bookmarksopen=false,%
colorlinks=true,%
linkcolor=blue,
citecolor=blue,
urlcolor=blue,
}

\renewcommand{\selectlanguage}[1]{} % to avoid the error of bibtex: language="en"

\begin{document}

\preprint{APS/123-QED}

\title{Collective excitations in active solids featuring alignment interactions}% Force line breaks with \\

\author{Yutaka Kinoshita}
 \email{yutaka.kinoshita.r4@dc.tohoku.ac.jp}
\affiliation{
 Institut f\"ur Physik, Otto-von-Guericke-Universit\"at Magdeburg, Universit\"atsplatz 2, 39106 Magdeburg, Germany
}%
\affiliation{
 Department of Physics, Tohoku University, Sendai 980-8578, Japan
}%
\author{Nariya Uchida}
\affiliation{
 Department of Physics, Tohoku University, Sendai 980-8578, Japan
}%
\author{Andreas M. Menzel}
 \email{a.menzel@ovgu.de}
\affiliation{
 Institut f\"ur Physik, Otto-von-Guericke-Universit\"at Magdeburg, Universit\"atsplatz 2, 39106 Magdeburg, Germany
}%

\date{\today}% It is always \today, today,
             %  but any date may be explicitly specified

\begin{abstract}
With increasing emphasis on the study of active solids, the features of these classes of nonequilibrium systems and materials beyond their mere existence shift into focus. One concept of active solids addresses them as active, self-propelled units that are elastically linked to each other. The emergence of orientationally ordered, collectively moving states in such systems has been demonstrated. We here analyze the excitability of such collectively moving elastic states. To this end, we determine corresponding fluctuation spectra. They indicate that collectively excitable modes exist in the migrating solid. Differences arise when compared to those of corresponding passive solids. We provide evidence that the modes of excitation associated with the intrinsic fluctuations are related to corresponding modes of entropy production. Overall, we hope to stimulate by our investigation future experimental studies that focus on excitations in active solids. 
\end{abstract}

%\keywords{Suggested keywords}%Use showkeys class option if keyword
                              %display desired
\maketitle

% \tableofcontents

\section{\label{sec:intro} Introduction}
Recently, the properties of elastic active matter have been studied increasingly, following intense previous investigations on active  fluids~\cite{Vicsek1995-wv,Vicsek2012-eu,Toner1995-ew,Bricard2013-yu,Marchetti2013-av,Peshkov2014-gv,reinken2024vortex}. Active solids comprise a broad variety of different nonequilibrium systems, for instance, biopolymeric networks~\cite{Koenderink2009-uu}, biological tissue~\cite{Peyret2019-tb,Armon2021-yy,Giavazzi2018-nb}, or biofilms~\cite{Xu2022-to,Chen2017-rw,Nijjer2022-ff}. 
Here, we refer to model systems in an elastically deformable state that are composed of active self-propelled objects~\cite{Hawkins2014-ff,Geyer2019-fj,Gorbushin2020-rm,Alonso2017-mn,Wu2023-uv,Banerjee2011-uo,Lin2021-ix,Turgut2020-nz,Menzel2013-ta,Woodhouse2018-gp,Briand2018-mi,Scheibner2020-di,Ophaus2021-da}.
Experimentally investigated examples of this kind include self-driven toys like Hexbugs$^\text{\textregistered}$ connected by elastically deformable springs~\cite{Baconnier2022-lf,Baconnier2023-qg}. Related phenomena comprise active jamming~\cite{Behringer2019-cf,Henkes2011-qc} and the crystallization of vibrated disks~\cite{Briand2016-db}.

As a consequence of the energy input into the system by each object, properties far beyond those of passive solids arise. 
For example, corresponding systems can become more rigid than those composed of passive Brownian particles when nonlinear elasticity and self-propulsion of the particles are combined~\cite{Sandoval2023-kr}. 
In addition, collective migration of solid-like active systems can arise~\cite{Menzel2012-up, Ferrante2013-mr, Menzel2014-cv, Menzel2013-ta}. Interestingly, the onset of collective motion was observed even in the absence of an explicit alignment mechanics of the individual propulsion directions \cite{Menzel2013-ta, Menzel2014-cv, Alaimo2016-in, Ophaus2018-wv, Ophaus2021-da}. 

Here, we focus on active solids that introduce aligning torques. They reorient the directions of self-propulsion of each elastically coupled object towards the current direction of the elastic force acting on this object. 
There are two obvious states in a bulk system that should be considered: orientationally disordered states concerning the directions of self-propulsion of each object and states of a certain degree of global alignment of these directions~\cite{Ferrante2013-fq,Ferrante2013-mr,Huang2021-js}.
Depending on the strengths of the alignment interactions, globally aligned states are conceivable, at least in finite-sized systems. 
Such global alignment is also observed for fluid systems featuring elastic collisions~\cite{Lam2015-av}.
Conversely, strong rotational diffusion may still destabilize the orientational order and lead to a disordered state~\cite{Ferrante2013-mr}.
Confinements can favor additional modes of collective motions such as chiral oscillations~\cite{Baconnier2022-lf,Baconnier2023-qg}.
They result from restrictions that the confinement imposes onto collective motion. 

While the static properties of active solids with alignment interaction have been studied~\cite{Huang2021-js}, the dynamical ones are still under development. Previous works focused on finite-sized systems~\cite{Baconnier2022-lf, Baconnier2023-qg, Baconnier2024-ft}.
Besides, it is reported that oscillations accompanied by entropy production arise when a solid state is formed by the accumulation of active Brownian particles~\cite{Caprini2020-fy,Caprini2021-ad,Caprini2023-ro}.
Since solid states are formed by a variety of active particles~\cite{Weber2014-vb, Yang2024-zc}, a general analysis is desirable. 

In our work, we analytically evaluate the dynamical properties of bulk systems in an orientationally ordered, collectively migrating state.
To this end, the spectra of fluctuations are calculated analytically. 
Besides, we analytically evaluate an expression for the rate of entropy production that follows from the conventional path-integral theory~\cite{Seifert2012-ri}.
One of the prominent characteristics of active matter is continuous energy dissipation accompanied by a positive rate of entropy production~\cite{Mandal2017-tm}.
This value is useful for quantifying how far away from the equilibrium the states are.
It appears that the coupling between self-propulsion and self-alignment of the migrating, elastically coupled objects introduces additional excitations that are absent in a corresponding passive system and accompanied by additional entropy production. 
Numerical simulations are performed for verification. 

The remaining parts are structured as follows. 
In Sec.~\ref{sec:model}, we introduce the model for active Brownian objects and the alignment interaction resulting from the action of the connecting elastic springs.
We derive the fluctuation spectrum in the moving state from the linearized equations of motion in Sec.~\ref{sec:structure}.
Next, we calculate the rate of entropy production in Sec.~\ref{sec:entropy} based on the results obtained in Sec.~\ref{sec:structure}.
Afterwards, in Sec.~\ref{sec:numeric}, we report our results from numerical simulations of the original equations of motion introduced in Sec.~\ref{sec:model}, which match well with the analytical solutions.
Finally, our conclusions are stated in Sec.~\ref{sec:conclusion}.

\section{\label{sec:model} Theoretical description}

\begin{figure}
    \centering
    \includegraphics{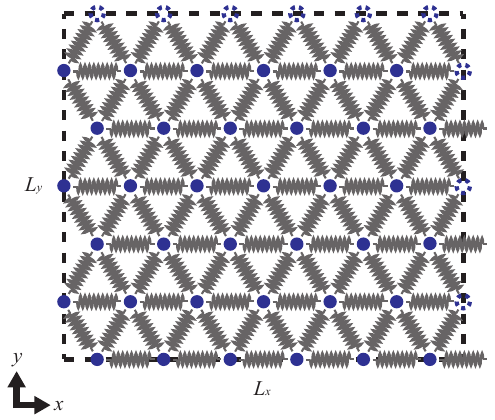}
    \caption{Schematic illustration of the system in its undeformed rest state. We impose periodic boundary conditions at each boundary. Colored objects are connected by springs. White objects are identical to the ones on the opposite side of the calculation box due to the boundary conditions.}
    \label{fig:system-schematic}
\end{figure}

To set up the active elastic solid, we consider $N$ identical active Brownian objects, each of mass $m$, pairwise connected by linear springs to form a two-dimensional hexagonal lattice, see Fig~\ref{fig:system-schematic}. 
The time evolution of the position $\boldsymbol{r}_i(t)$  and velocity $\boldsymbol{v}_i(t)$ of the $i$th object ($i=1,...,N$) is quantified by the translational equations of motion
\begin{align}
    \label{eq:coordinate}\dot{\boldsymbol{r}}_i(t)=& {}\;\boldsymbol{v}_i(t),\\
    \label{eq:velocity}m\dot{\boldsymbol{v}}_i(t)=& {} -\bm{\gamma}_r\boldsymbol{v}_i(t)-\boldsymbol{\nabla}_{\boldsymbol{r}_i}U(\left\{\boldsymbol{r}\right\})+F_0\boldsymbol{e}_i\nonumber\\
    &+\sqrt{2\gamma_r k_B T_r}\boldsymbol{\eta}_{i}(t).
\end{align}
In these expressions, dots mark time derivatives, and ${\gamma}_r$ is the coefficient of translational friction.
$T_r$ is an effective temperature and $k_B$ represents Boltzmann's constant. 
$U(\left\{\boldsymbol{r}\right\})$ includes the overall potential due to the Hookean springs introduced between neighboring objects in the undeformed ground state of the lattice. We assume periodic boundary conditions so that each object is permanently linked by elastic springs to its six surrounding initially nearest neighbors in our two-dimensional hexagonal lattice. 
$\left\{\boldsymbol{r}\right\}$ is a short notation for the tuple of all positional vectors $\boldsymbol{r}_i$ ($i=1,...,N$). 
In our case, the potential reads 
\begin{equation}
        U(\left\{\boldsymbol{r}\right\}) = \frac{1}{2}K\sum_{(i,j)}
        \left( \left|\boldsymbol{r}_i-\boldsymbol{r}_j\right|-a\right)^2.
        \label{eq:U}
\end{equation}
Here, $K$ quantifies the identical elastic stiffness of each spring, $a$ denotes the natural length of each spring in the ground state, and $(i,j)$ denotes all pairs of objects permanently linked by springs, without double-counting.
Moreover, each object self-propels with the identical strength of active driving $F_0$ into the individual direction of the active driving force $\boldsymbol{e}_i=\left(\cos\theta_i,\sin\theta_i\right)^T$, where $^T$ marks the transpose. $\theta_i$ is an angle in the two-dimensional $x$-$y$-plane measured from the $x$-axis. 
Lastly, $\boldsymbol{\eta}_{i}(t)$ represents the influence of a Gaussian-distributed white stochastic force. It has zero mean and satisfies $\braket{\eta_{i,\alpha}(t)\eta_{j,\beta}(t^\prime)}=\delta_{ij}\delta_{\alpha\beta}\delta(t-t^\prime)$, $\alpha,\beta\in \{x,y\}$ in two dimensions. $\delta_{ij}$ marks the Kronecker delta and $\delta(t-t')$ the Dirac delta function. 

Equations~\eqref{eq:coordinate} and \eqref{eq:velocity} for translational motion are supplemented by the equations of rotational motion for the direction of self-propulsion $\boldsymbol{e}_i$ of each object $i$ ($i=1, ... , N$). Introducing the angular frequency $\omega_i$ of each object $i$, these equations read
\begin{align}
    \label{eq:rotin}\dot{\theta}_i=&\;\omega_i,\\
    \label{eq:theta}I\dot{\omega}_i=&{}-\gamma_\theta\omega_i+
    \zeta|\boldsymbol{v}_i|\sin(\phi_i-\theta_i) \nonumber\\
    &{}+\sqrt{2\gamma_\theta k_B T_\theta}\,\eta_{\theta,i}(t).
\end{align}
Here, $I$ plays the role of the identical moment of inertia of each object, $\gamma_\theta$ is an identical coefficient of rotational friction of each object, while $\zeta$ sets the identical strength of orientational alignment between the orientational angle $\theta_i$ of the direction of self-propulsion $\boldsymbol{e}_i$ of the object $i$ and the orientational angle $\phi_i$ of the velocity \cite{Baconnier2022-lf, Baconnier2023-qg}. $\phi_i$ is not only affected by active driving but also by the elastic forces.
We consider $\zeta$ to be positive, which means the direction of self-driving tends to orient along the direction of the elastic force. 
$\eta_{\theta,i}(t)$ includes the effect of a white Gaussian stochastic torque with the effective temperature $T_{\theta}$. Its mean vanishes and it satisfies $\braket{\eta_{\theta,i}(t)\eta_{\theta,j}(t^\prime)}=\delta_{ij}\delta(t-t^\prime)$.

We rescale lengths according to $\boldsymbol{r}=a\hat{\boldsymbol{r}}$ and times according to $t=\gamma_r\hat{t}/K$. Hats denote dimensionless quantities. Thus, we arrive at the dimensionless coupled equations of motion
\begin{align}
    \label{eq:dimless:r}\dot{\hat{\boldsymbol{r}}}_i=&\;\hat{\boldsymbol{v}}_i,\\[.1cm]
    \label{eq:dimless:v}\tau_v\dot{\hat{\boldsymbol{v}}}_i=&{}-\hat{\boldsymbol{v}}_i-\sum_%{j\in L_i}
    {j\in (i,j)}\left(|\hat{\boldsymbol{r}}_i-\hat{\boldsymbol{r}}_j|-1\right)\frac{\hat{\boldsymbol{r}}_i-\hat{\boldsymbol{r}}_j}{|\hat{\boldsymbol{r}}_i-\hat{\boldsymbol{r}}_j|} \nonumber\\
    &{}+\alpha\boldsymbol{e}_i+\sqrt{2D_r}\hat{\boldsymbol{\eta}}_i(\hat{t}), \\
    \label{eq:dimless:theta}\dot{\theta}_i=&\;\hat{\omega}_i,\\[.1cm]    
    \label{eq:dimless:omega}\tau_\theta\dot{\hat{\omega}}_i=&{}-\hat{\omega}_i+\xi\hat{v}_i\sin(\phi_i-\theta_i)+\sqrt{2D_\theta}\hat{\eta}_{\theta,i}(\hat{t}).
\end{align}
In these expressions, we define $\tau_v=mK/\gamma_r^2$, $D_r=k_BT_r/a^2K$, $\alpha=F_0/Ka$, $\tau_\theta=IK/%b
\gamma_r\gamma_\theta$, $\xi=\zeta a/\gamma_\theta$, and $D_\theta=k_BT_\theta \gamma_r/K\gamma_\theta$.

$\hat{\eta}_{\theta,i}(t)$ is a white Gaussian stochastic process of zero mean and unit variance.
Moreover, we omit the hat to simplify notation.
We set $\tau_v=\tau_\theta=0$ because we only consider the overdamped situation for simplicity, in line with experiments and considerations on a colloidal scale. 

\section{\label{sec:structure} Fluctuation spectra in the collectively migrating state}
As demonstrated earlier~\cite{Huang2021-js}, the system shows an ordered state of all objects moving collectively on average in the same direction, while maintaining a hexagonal lattice structure. 
Let us denote the angle of the global direction of migration as $\theta_0$ and the overall speed of the lattice structure as $V$.
We relate $V$ and $\theta_0$ to the polar order parameter
\begin{equation}
    \boldsymbol{P}=\frac{1}{N}\sum_i^N\boldsymbol{e}(\theta_i)
    \label{eq:polar-order}
\end{equation}
through the equation $V\boldsymbol{e}(\theta_0)=\alpha\boldsymbol{P}$.
For vanishing noise, the global speed $V$ in the perfectly ordered state matches the speed of self-propulsion of each object. In this case, $V=\alpha$ because $|\boldsymbol{P}|=1$.
The speed $V$ decreases with increasing diffusion constants, accompanied by a decreasing degree of polar orientational order $|\boldsymbol{P}|$.
According to the large-number theorem, the fluctuations of both the speed and the angle of collective motion tend to zero in the limit of large system sizes.
In this limit, the global velocity of the system is the same as the averaged velocity of each object, $\left\langle\dot{\boldsymbol{r}_i}\right\rangle=V\boldsymbol{e}(\theta_0)$ for all $i=1,...,N$.
Similarly, we register for the angles $\theta_i$ that $\braket{\theta_i}=\theta_0$.

Next, we denote each reference position as $\boldsymbol{x}^{(0)}_i(t)=(x^{(0)}_i(t),y^{(0)}_i(t))^T$ for $i=1,..., N$. It is obtained from the starting position in the perfect hexagonal lattice structure $\boldsymbol{x}^{(0)}_i(t=0)=\boldsymbol{x}^{(0)}_i$ via $\boldsymbol{x}^{(0)}_i(t)=\boldsymbol{x}^{(0)}_i+V\boldsymbol{e}(\theta_0)t$. 

We assume that the deviations from this idealized situation induced by stochastic fluctuations as quantified by the diffusion constants are small. 
In other words, the diffusion constants $D_r$ and $D_\theta$ are low in magnitude.
To this end, we introduce the displacements $\boldsymbol{u}_i(t)=\boldsymbol{r}_i(t)-\boldsymbol{x}_i^{(0)}(t)$ and the deviations in the individual angles of self-propulsion $\delta\theta_i(t)=\theta_i(t)-\theta_0$. Therefore, 
\begin{subequations}
    \begin{equation}
        \boldsymbol{r}_i(t)=
        \boldsymbol{x}^{(0)}_i+V\boldsymbol{e}(\theta_0) t+ \boldsymbol{u}_i(t),
    \end{equation}
\begin{equation}
    \theta_i(t)=
    \theta_0 + \delta\theta_i(t).
\end{equation}
\end{subequations}

The elastic potential $\hat{U}(\{\boldsymbol{u}\})$, where $\{\boldsymbol{u}\}$ summarizes all $\boldsymbol{u}_i$ ($i=1,...,N$) and $\boldsymbol{x}_{ij}^{(0)}=\boldsymbol{x}_i^{(0)}-\boldsymbol{x}_j^{(0)} $, is approximated to second order in $\boldsymbol{u}_{ij}=\boldsymbol{u}_i-\boldsymbol{u}_j $,
\begin{equation}
\begin{split}
    \hat{U}(\{\boldsymbol{u}\})&=\frac{1}{2}\sum_{(i,j)}
    \left(|\boldsymbol{x}^{(0)}_{ij}+\boldsymbol{u}_{ij}|-1\right)^2\\
    &\approx \frac{1}{2}\sum_{(i,j)}
    \left(\boldsymbol{x}^{(0)}_{ij}\cdot\boldsymbol{u}_{ij}\right)^2.
\end{split}
\end{equation}
Thus, the elastic forces are approximated to first order in $\boldsymbol{u}$ as
\begin{equation}
    \boldsymbol{F}_i\approx{}-\sum_{j\in (i,j)}\left( \boldsymbol{x}_{ij}^{(0)}\cdot\boldsymbol{u}_{ij} \right)\boldsymbol{x}_{ij}^{(0)}={}-\sum_{j\in (i,j)}\underline{\boldsymbol{M}}_{ij}\cdot\boldsymbol{u}_{ij},
    \label{eq:elasticforce:realspace}
\end{equation}
where, together with the parameterization $\boldsymbol{x}^{(0)}_{ij}=\left(x^{(0)}_{ij},y^{(0)}_{ij}\right)^T$, the matrix $\underline{\boldsymbol{M}}_{ij}$ is given by 
\begin{equation}
    \underline{\boldsymbol{M}}_{ij}=
    \begin{pmatrix}
        \left(x_{ij}^{(0)}\right)^2 & x_{ij}^{(0)} y_{ij}^{(0)}\\[.1cm]
        x_{ij}^{(0)} y_{ij}^{(0)} & \left(y_{ij}^{(0)}\right)^2.
    \end{pmatrix}.
    \label{eq:elastic:realspace}
\end{equation}

Using these abbreviations,
the linearized equation for translational motion following from Eq.~(\ref{eq:dimless:v}) becomes
\begin{align}
        \dot{\boldsymbol{u}}_i=&\,{}-\sum_{j\in (i,j)}\underline{\boldsymbol{M}}_{ij}\cdot\boldsymbol{u}_{ij}+\left(\alpha-V\right)\boldsymbol{e}(\theta_0)
        \nonumber\\
        &\,{}+\alpha\begin{pmatrix}-\sin\theta_0 \\ \cos\theta_0\end{pmatrix}\delta\theta_i+\boldsymbol{\eta}_i(t).
        \label{eq:V-alpha}
\end{align}
We may conclude that this equation simplifies in the limit of large numbers of linked objects $N$.
For demonstration, we expand the polar orientational order parameter as
\begin{equation}
    \begin{split}
        \boldsymbol{P}=&\frac{1}{N}\sum_{i=1}^N\boldsymbol{e}\left(\theta_0+\delta\theta_i\right)\\
        \approx&\frac{1}{N}\sum_{i=1}^N\left(\boldsymbol{e}\left(\theta_0\right)+\boldsymbol{e}^\prime\!\left(\theta_0\right)\delta\theta_i\right)\\
        =&\frac{1}{N}\sum_{i=1}^N\boldsymbol{e}\left(\theta_0\right)+\boldsymbol{e}^\prime\!\left(\theta_0\right)\,\frac{1}{N}\sum_{i=1}^N\delta\theta_i
        \nonumber\\
        =&\,1+\boldsymbol{e}^\prime\!\left(\theta_0\right)\,\frac{1}{N}\sum_{i=1}^N\delta\theta_i.
    \end{split}
\end{equation}
The second term of the last expression becomes zero in the limit of large numbers of linked objects $N\rightarrow\infty$.
Then, assuming that the globally orientationally ordered state exists, $\boldsymbol{P}$ satisfies the equality $|\boldsymbol{P}|=1$ to linear order.
We numerically confirm the validity of this approximation for the considered scenario in Sec~\ref{sec:numeric}. 
As a consequence, $V\boldsymbol{e}(\theta_0)=\alpha\boldsymbol{P}$ for $N\rightarrow\infty$ implies $V=\alpha$. Accordingly, the second term on the right-hand side of Eq.~\eqref{eq:V-alpha} becomes zero.

Thus, the coupled linearized equations of motion are derived from Eqs.~(\ref{eq:dimless:v}) and (\ref{eq:dimless:omega}) as
    \begin{align}
        \dot{\boldsymbol{u}}_i=&{}-\sum_{j\in (i,j)}\underline{\boldsymbol{M}}_{ij}\cdot\boldsymbol{u}_{ij}+\alpha\begin{pmatrix}-\sin\theta_0 \\ \cos\theta_0\end{pmatrix}\delta\theta_i
        \nonumber\\[.1cm]&{}+\sqrt{2D_r}\boldsymbol{\eta}_i(t),
        \label{eq:u-dot}\\[.2cm]
        \dot{\delta\theta}_i=&\;-\xi\sum_{j\in (i,j)}\left[\cos\theta_0 (\underline{\boldsymbol{M}}_{ij}\cdot\boldsymbol{u}_{ij})_y-\sin\theta_0 (\underline{\boldsymbol{M}}_{ij}\cdot\boldsymbol{u}_{ij})_x\right]\nonumber\\
        &{}-\xi\sin\theta_0\sqrt{2D_r}\eta_{x,i}(t)+\xi\cos\theta_0\sqrt{2D_r}\eta_{y,i}(t)
        \nonumber\\[.1cm]&
        {}+\sqrt{2D_\theta}\eta_{\theta,i}(t).
        \label{eq:delta-theta-dot}
    \end{align}
We apply the Fourier series in space and combine Eqs.~(\ref{eq:u-dot}) and (\ref{eq:delta-theta-dot}) into one equation as
\begin{equation}
    \begin{split}
        \dot{\boldsymbol{y}}_{\boldsymbol{k}} =& -\underline{\boldsymbol{\Gamma}}^{(xy)}_{\boldsymbol{k}}\cdot \boldsymbol{y}_{\boldsymbol{k}}+\underline{\boldsymbol{D}}^{(xy)}\cdot\boldsymbol{\eta}_{\boldsymbol{k}}(t).
    \label{eq:cartesian}
    \end{split}
\end{equation}
Here, the stochastic variables are combined into one vector $\boldsymbol{y}_{\boldsymbol{k}}=(u_{x,\boldsymbol{k}}, u_{y,\boldsymbol{k}}, \delta\theta_{\boldsymbol{k}})^T$. 
In Appendix~\ref{appendix:fourier}, we include the conventions that we employ for Fourier transformation in time and Fourier series in space.
The coefficients on the right-hand sides of the equations are summarized by the matrices\\
\begin{widetext}
\begin{equation}
    \underline{\boldsymbol{\Gamma}}_{\boldsymbol{k}}^{(xy)}=
    \begin{pmatrix}
        M_{xx,\boldsymbol{k}} & M_{xy,\boldsymbol{k}} & \alpha\sin\theta_0 \\
        M_{yx,\boldsymbol{k}} & M_{yy,\boldsymbol{k}}& -\alpha\cos\theta_0 \\
        \xi(M_{yx,\boldsymbol{k}}\cos\theta_0-M_{xx,\boldsymbol{k}}\sin\theta_0) & \xi(M_{yy,\boldsymbol{k}}\cos\theta_0-M_{xy,\boldsymbol{k}}\sin\theta_0) & 0
    \end{pmatrix},
\end{equation}
\end{widetext}
and 
\begin{equation}
    \underline{\boldsymbol{D}}^{(xy)}=
    \begin{pmatrix}
        \sqrt{2D_r} & 0 & 0 \\
        0 & \sqrt{2D_r} & 0 \\
        -\xi\sin\theta_0 \sqrt{2D_r} & \xi\cos\theta_0\sqrt{2D_r} & \sqrt{2D_\theta}
    \end{pmatrix}.
\end{equation}
We define these expressions $\boldsymbol{\eta}_{\boldsymbol{k}}(t)=(\eta_{x,\boldsymbol{k}}(t), \eta_{y,\boldsymbol{k}}(t), \eta_{\theta, \boldsymbol{k}}(t))^T$ as the vector of the stochastic contributions.
The definition of the matrix  $\underline{\boldsymbol{M}}_{\boldsymbol{k}}$ is elucidated in Appendix~\ref{appendix:elastic}.
Explicitly, its components read
\begin{equation}
    \underline{\boldsymbol{M}}_{\boldsymbol{k}}=
    \begin{pmatrix}
        M_{xx, \boldsymbol{k}} & M_{xy, \boldsymbol{k}} \\
        M_{xy, \boldsymbol{k}} & M_{yy, \boldsymbol{k}} \\
    \end{pmatrix},
    \label{eq:elastic_k}
\end{equation}
where
\begin{widetext}
\begin{subequations}
\begin{equation}
        M_{xx,\boldsymbol{k}}=3-2\cos(k_x)-\frac{1}{2}\cos\left(\frac{1}{2}k_x+\frac{\sqrt{3}}{2}k_y\right)-\frac{1}{2}\cos\left(\frac{1}{2}k_x-\frac{\sqrt{3}}{2}k_y\right)
\end{equation}
\begin{equation}
        M_{xy,\boldsymbol{k}}=-\frac{\sqrt{3}}{2}\cos\left(\frac{1}{2}k_x+\frac{\sqrt{3}}{2}k_y\right)+\frac{\sqrt{3}}{2}\cos\left(\frac{1}{2}k_x-\frac{\sqrt{3}}{2}k_y\right)
\end{equation}
\begin{equation}
        M_{yy,\boldsymbol{k}}=3-\frac{3}{2}\cos\left(\frac{1}{2}k_x+\frac{\sqrt{3}}{2}k_y\right)-\frac{3}{2}\cos\left(\frac{1}{2}k_x-\frac{\sqrt{3}}{2}k_y\right)
\end{equation}
\end{subequations}
\end{widetext}
From the symmetry of the cosine function, we infer $\underline{\boldsymbol{M}}_{\boldsymbol{k}}=\underline{\boldsymbol{M}}_{-\boldsymbol{k}}$, which reflects the inversion symmetry of the system. In addition, $\underline{\boldsymbol{M}}_{\boldsymbol{k}}$ itself is symmetric.

To simplify our considerations, we rotate the coordinate frame in Eq.~\eqref{eq:cartesian} by $\theta_0$.
The rotation matrix is defined by
\begin{equation}
    \underline{\boldsymbol{R}}(\theta)=
    \begin{pmatrix}
        \cos\theta & -\sin\theta & 0 \\
        \sin\theta & \cos\theta & 0 \\
        0 & 0 & 1
    \end{pmatrix}.
\end{equation}

Through rotation by $\theta_0$ in anticlockwise direction, we obtain $\boldsymbol{z}_{\boldsymbol{k}}=(u_{\parallel,\boldsymbol{k}}, u_{\perp,\boldsymbol{k}}, \delta\theta_{\boldsymbol{k}})^T$ and $\boldsymbol{\eta}^{(\theta_0)}_{\boldsymbol{k}}=(\eta_{\parallel,\boldsymbol{k}}(t), \eta_{\perp,\boldsymbol{k}}(t), \eta_{\theta,\boldsymbol{k}}(t))^T$ via
\begin{equation}
        \boldsymbol{y}_{\boldsymbol{k}} = \underline{\boldsymbol{R}}(\theta_0)\cdot\boldsymbol{z}_{\boldsymbol{k}}, \qquad
        \boldsymbol{\eta}_{\boldsymbol{k}} = \underline{\boldsymbol{R}}(\theta_0)\cdot\boldsymbol{\eta}^{(\theta_0)}_{\boldsymbol{k}}. 
\end{equation}
Hereafter, we set $\boldsymbol{\eta}_{\boldsymbol{k}}(t)\equiv\boldsymbol{\eta}^{(\theta_0)}_{\boldsymbol{k}}(t)$, because the properties are the same.
Our equation in the rotated coordinate frame becomes
\begin{equation}
    \begin{split}
        \dot{\boldsymbol{z}}_{\boldsymbol{k}}
        + \underline{\boldsymbol{\Gamma}}^{(\theta_0)}_{\boldsymbol{k}}
        \cdot \boldsymbol{z}_{\boldsymbol{k}}
        =\underline{\boldsymbol{D}}^{(\theta_0)}\cdot\boldsymbol{\eta}_{\boldsymbol{k}}(t),
    \label{eq:rotated}
    \end{split}
\end{equation}
where
\begin{equation}
\begin{split}
    \underline{\boldsymbol{D}}^{(\theta_0)}=&\underline{\boldsymbol{R}}(-\theta_0)\cdot\underline{\boldsymbol{D}}^{(xy)}\cdot\underline{\boldsymbol{R}}(\theta_0)\\
    =&
    \begin{pmatrix}
        \sqrt{2D_r} & 0 & 0 \\
        0 & \sqrt{2D_r} & 0 \\
        0 & \xi\sqrt{2D_r} & \sqrt{2D_\theta}
    \end{pmatrix}
    \label{eq:diffusion-rot}
\end{split}
\end{equation}
and
\begin{equation}
    \begin{split}
    \underline{\boldsymbol{\Gamma}}_{\boldsymbol{k}}^{(\theta_0)}=&\underline{\boldsymbol{R}}(-\theta_0)\cdot\underline{\boldsymbol{\Gamma}}^{(xy)}_{\boldsymbol{k}}\cdot\underline{\boldsymbol{R}}(\theta_0)\\
    =&
    \begin{pmatrix}
        A_{\boldsymbol{k}}+B_{\boldsymbol{k}} & C_{\boldsymbol{k}} & 0 \\
        C_{\boldsymbol{k}} & A_{\boldsymbol{k}}-B_{\boldsymbol{k}} & -\alpha \\
        \xi C_{\boldsymbol{k}} & \xi\left(A_{\boldsymbol{k}}-B_{\boldsymbol{k}}\right) & 0
    \end{pmatrix}.
    \label{eq:gamma-rot}
\end{split}
\end{equation}
In this way, we introduced the abbreviations 
\begin{subequations}
\begin{equation}
        A_{\boldsymbol{k}}=\frac{M_{xx,\boldsymbol{k}}+M_{yy,\boldsymbol{k}}}{2},
\end{equation}
\begin{equation}
        B_{\boldsymbol{k}}=\frac{M_{xx,\boldsymbol{k}}-M_{yy,\boldsymbol{k}}}{2}\cos2\theta_0 + M_{xy,\boldsymbol{k}}\sin2\theta_0,
\end{equation}
\begin{equation}
        C_{\boldsymbol{k}}=-\frac{M_{xx,\boldsymbol{k}}-M_{yy,\boldsymbol{k}}}{2}\sin2\theta_0 + M_{xy,\boldsymbol{k}}\cos2\theta_0.
\end{equation}
\end{subequations}

We solve Eq.~\eqref{eq:rotated} using the method of Green's functions~\cite{Loos2020-jt}.
Firstly, we perform the Fourier transformation in time, and the formal solution $\tilde{\boldsymbol{z}}_{\boldsymbol{k}}(\omega)=\mathcal{F}\left[\boldsymbol{z}_{\boldsymbol{k}}(t)\right]$ is expressed via
\begin{equation}
        -i\omega\tilde{\boldsymbol{z}}_{\boldsymbol{k}}
        + \underline{\boldsymbol{\Gamma}}^{(\theta_0)}_{\boldsymbol{k}}
        \cdot \tilde{\boldsymbol{z}}_{\boldsymbol{k}}
=\underline{\boldsymbol{D}}^{(\theta_0)}\cdot\tilde{\boldsymbol{\eta}}_{\boldsymbol{k}}(\omega)
\label{eq_equationofmotion}
\end{equation}
as
\begin{equation}
\tilde{\boldsymbol{z}}_{\boldsymbol{k}}(\omega)=\left[-i\omega\underline{\boldsymbol{I}}+\underline{\boldsymbol{\Gamma}}^{(\theta_0)}_{\boldsymbol{k}} \right]^{-1}\cdot \underline{\boldsymbol{D}}^{(\theta_0)}\cdot\tilde{\boldsymbol{\eta}}_{\boldsymbol{k}}(\omega).
\label{eq:formal-solution}
\end{equation}
The stochastic contributions in Fourier space satisfy the relations
\begin{subequations}
\begin{equation}
        \braket{\tilde{\boldsymbol{\eta}}_{\boldsymbol{k}}(\omega)}=\boldsymbol{0}
\end{equation}
and
\begin{equation}
        \braket{\tilde{\eta}_{\alpha,\boldsymbol{k}}(\omega)\tilde{\eta}_{\beta,\boldsymbol{k}^\prime}(\omega^\prime)}=2\pi N\delta_{\alpha,\beta}\delta_{\boldsymbol{k},-\boldsymbol{k}^\prime}\delta(\omega+\omega^\prime),
\end{equation}
\begin{equation}
        \braket{\tilde{\eta}_{\alpha,\boldsymbol{k}}(\omega)\tilde{\eta}^\ast_{\beta,\boldsymbol{k}^\prime}(\omega^\prime)}=2\pi N\delta_{\alpha,\beta}\delta_{\boldsymbol{k},\boldsymbol{k}^\prime}\delta(\omega-\omega^\prime)
\end{equation}
\label{eq:noise_fourier}
\end{subequations}
for $\alpha,\beta\in\{x,y,\theta\}$.

Next, introducing the abbreviation 
\begin{equation}
    \underline{\boldsymbol{\lambda}}_{\boldsymbol{k}}(\omega)\coloneq \left[-i\omega\underline{\boldsymbol{I}}+\underline{\boldsymbol{\Gamma}}^{(\theta_0)}_{\boldsymbol{k}} \right]^{-1},
\end{equation}
the autocorrelation matrix is defined as
\begin{widetext}
\begin{equation}
  \begin{split}
    \underline{\boldsymbol{S}}_{\boldsymbol{k},\boldsymbol{k}^\prime}(\omega,\omega^\prime)\coloneq\braket{\tilde{\boldsymbol{z}}_{\boldsymbol{k}}(\omega)\otimes\tilde{\boldsymbol{z}}^{\ast}_{\boldsymbol{k}^\prime}(\omega^\prime)}=&
    \begin{pmatrix}
        \braket{\tilde{u}_{\parallel,\boldsymbol{k}}(\omega)\tilde{u}^\ast_{\parallel,\boldsymbol{k}^\prime}(\omega^\prime)} & \braket{\tilde{u}_{\parallel,\boldsymbol{k}}(\omega)\tilde{u}^\ast_{\perp,\boldsymbol{k}^\prime}(\omega^\prime)} & \braket{\tilde{u}_{\parallel,\boldsymbol{k}}(\omega)\tilde{\theta}^\ast_{\boldsymbol{k}^\prime}(\omega^\prime)} \\
 \braket{\tilde{u}_{\perp,\boldsymbol{k}}(\omega)\tilde{u}^\ast_{\parallel,\boldsymbol{k}^\prime}(\omega^\prime)} & \braket{\tilde{u}_{\perp,\boldsymbol{k}}(\omega)\tilde{u}^\ast_{\perp,\boldsymbol{k}^\prime}(\omega^\prime)} & \braket{\tilde{u}_{\perp,\boldsymbol{k}}(\omega)\tilde{\theta}^\ast_{\boldsymbol{k}^\prime}(\omega^\prime)} \\
 \braket{\tilde{\theta}_{\boldsymbol{k}}(\omega)\tilde{u}^\ast_{\parallel,\boldsymbol{k}^\prime}(\omega^\prime)} & \braket{\tilde{\theta}_{\boldsymbol{k}}(\omega)\tilde{u}^\ast_{\perp,\boldsymbol{k}^\prime}(\omega^\prime)} & \braket{\tilde{\theta}_{\boldsymbol{k}}(\omega)\tilde{\theta}^\ast_{\boldsymbol{k}^\prime}(\omega^\prime)}
    \end{pmatrix}
    \end{split}
\end{equation}
\end{widetext}
where $\otimes$ denotes the dyadic product. 
We may rewrite this expression as 
\begin{align}
    \underline{\tilde{\boldsymbol{S}}}_{\boldsymbol{k},\boldsymbol{k}'}(\omega,\omega')
    &=
    \braket{\tilde{\boldsymbol{z}}_{\boldsymbol{k}}(\omega)\otimes\tilde{\boldsymbol{z}}^\ast_{\boldsymbol{k}'}(\omega')}\nonumber\\
    &=\braket{\tilde{\boldsymbol{z}}_{\boldsymbol{k}}(\omega)\otimes\tilde{\boldsymbol{z}}_{-\boldsymbol{k}'}(-\omega')} \nonumber\\
    &=2\pi N \underline{\boldsymbol{S}}_{\boldsymbol{k}}(\omega) 
    \,\delta_{\boldsymbol{k},{\boldsymbol{k}'}}
    \,\delta(\omega-\omega'),
    \label{eq:dynamical-structures}
\end{align}
where in the last step we introduced the matrix of equal-frequency and equal-wavevector fluctuation spectra $\underline{\boldsymbol{S}}_{\boldsymbol{k}}(\omega)$ with components
\begin{equation}
    \begin{split}
        S_{\alpha\beta,\boldsymbol{k}}(\omega)=\lambda_{\alpha \gamma,\boldsymbol{k}}(\omega)D_{\gamma \eta}^{(\theta_0)}D_{\eta\zeta}^{(\theta_0)T}\lambda^{T}_{\zeta\beta, \boldsymbol{k}}(-\omega). 
        \label{eq:S-k-omega}
    \end{split}
\end{equation}
To derive this equation, we used the relation $\underline{\boldsymbol{\lambda}}_{\boldsymbol{k}}(\omega)=\underline{\boldsymbol{\lambda}}_{-\boldsymbol{k}}(\omega)$, the formal solution Eq.~\eqref{eq:formal-solution}, and we applied Eqs.~\eqref{eq:noise_fourier}.
Explicitly, we list the components as
\begin{align}
        S_{\alpha\beta,\boldsymbol{k}}(\omega)=\frac{W^{(r)}_{\alpha\beta,\boldsymbol{k}}(\omega)2D_r+W^{(\theta)}_{\alpha\beta,\boldsymbol{k}}(\omega)2D_\theta}{(\mathrm{denom.})},
        \label{eq:dynamical_structure}
\end{align}
where we introduce the following abbreviations
\begin{widetext}
    \begin{equation}
        \begin{split}
            (\mathrm{denom.})=&\omega^2\left(\omega^2 + A_{\boldsymbol{k}}^2 + B_{\boldsymbol{k}}^2 + C_{\boldsymbol{k}}^2 + 2A_{\boldsymbol{k}}\sqrt{B_{\boldsymbol{k}}^2+C_{\boldsymbol{k}}^2} \right)\left(\omega^2 + A_{\boldsymbol{k}}^2 + B_{\boldsymbol{k}}^2 + C_{\boldsymbol{k}}^2 - 2A_{\boldsymbol{k}}\sqrt{B_{\boldsymbol{k}}^2+C_{\boldsymbol{k}}^2} \right)\\
            &-2\alpha\xi\omega^2\left[(A_{\boldsymbol{k}}-B_{\boldsymbol{k}})\omega^2 +(A_{\boldsymbol{k}}+B_{\boldsymbol{k}})(A_{\boldsymbol{k}}^2-B_{\boldsymbol{k}}^2-C_{\boldsymbol{k}}^2) \right]\\
            &+\alpha^2\xi^2\left[ (A_{\boldsymbol{k}}-B_{\boldsymbol{k}})^2\omega^2 +\left(A_{\boldsymbol{k}}^2-B_{\boldsymbol{k}}^2-C_{\boldsymbol{k}}^2\right)^2 \right],\\
        \end{split}
    \end{equation}
\begin{subequations}
        \begin{equation}
        W^{(r)}_{\parallel\parallel,\boldsymbol{k}}(\omega)=\omega^4 + \left( \left(A_{\boldsymbol{k}}-B_{\boldsymbol{k}}\right)\left(A_{\boldsymbol{k}}-B_{\boldsymbol{k}}-2\alpha\xi\right)+C_{\boldsymbol{k}}^2 \right)\omega^2+\alpha^2\xi^2\left( \left(A_{\boldsymbol{k}}-{B}_{\boldsymbol{k}}\right)^2+C_{\boldsymbol{k}}^2 \right),
        \end{equation}
        \begin{equation}
        W^{(r)}_{\perp\perp,\boldsymbol{k}}(\omega)=\left(\alpha^{2} \xi^{2} + \omega^{2}\right) \left(A_{\boldsymbol{k}}^{2} + 2 A_{\boldsymbol{k}} B_{\boldsymbol{k}} + B_{\boldsymbol{k}}^{2} + C_{\boldsymbol{k}}^{2} + \omega^{2}\right),\\
        \end{equation}
        \begin{equation}
        W^{(r)}_{\theta\theta,\boldsymbol{k}}(\omega)=\xi^2\omega^2 \left[ \omega^2+ (A_{\boldsymbol{k}}+B_{\boldsymbol{k}})^2+C_{\boldsymbol{k}}^2 \right],\\
        \end{equation}
        \begin{equation}
        W^{(r)}_{\perp\parallel,\boldsymbol{k}}(\omega)=C_{\boldsymbol{k}} \left[( \alpha\xi-2A_{\boldsymbol{k}} )\omega^2-2\alpha^2\xi^2 A_{\boldsymbol{k}} +i\alpha^2\xi^2 \omega \right],\\
        \end{equation}
        \begin{equation}
        W^{(r)}_{\theta \parallel,\boldsymbol{k}}(\omega)=C_{\boldsymbol{k}} \left(\left(\alpha\xi^2 -2\xi A_{\boldsymbol{k}} \right)\omega^2 + 2i\alpha A_{\boldsymbol{k}} \xi^2\omega\right),\\
        \end{equation}
        \begin{equation}
        W^{(r)}_{\theta \perp,\boldsymbol{k}}(\omega)=\left(\xi\omega^2-i\alpha\xi^2\omega\right)\left[\omega^2+\left(A_{\boldsymbol{k}}+B_{\boldsymbol{k}}\right)^2+C^2_{\boldsymbol{k}}\right],\\
        \end{equation}
        \begin{equation}
        W^{(\theta)}_{\parallel\parallel,\boldsymbol{k}}(\omega)=\alpha^2 C_{\boldsymbol{k}}^2,\\
        \end{equation}
        \begin{equation}
        W^{(\theta)}_{\perp\perp,\boldsymbol{k}}(\omega)=\alpha^2\left( \omega^2 +(A_{\boldsymbol{k}}+B_{\boldsymbol{k}})^2 \right),\\
        \end{equation}
        \begin{equation}
        W^{(\theta)}_{\theta\theta,\boldsymbol{k}}(\omega)= \omega^4+ 2(A_{\boldsymbol{k}}^2+B_{\boldsymbol{k}}^2+C_{\boldsymbol{k}}^2)\omega^2+(A_{\boldsymbol{k}}^2-B_{\boldsymbol{k}}^2-C_{\boldsymbol{k}}^2)^2,\\
        \end{equation}
        \begin{equation}
        W^{(\theta)}_{\perp\parallel,\boldsymbol{k}}(\omega)=- C_{\boldsymbol{k}} \alpha^{2} \left(A_{\boldsymbol{k}} + B_{\boldsymbol{k}} - i \omega\right),\\
        \end{equation}
        \begin{equation}
        W^{(\theta)}_{\theta \parallel,\boldsymbol{k}}(\omega)=- C_{\boldsymbol{k}} \alpha \left(A_{\boldsymbol{k}}^{2} - 2 i A_{\boldsymbol{k}} \omega - B_{\boldsymbol{k}}^{2} - C_{\boldsymbol{k}}^{2} - \omega^{2}\right),\\
        \end{equation}
        \begin{equation}
        W^{(\theta)}_{\theta \perp,\boldsymbol{k}}(\omega)=\alpha\left[ \omega^2\left(A_{\boldsymbol{k}}-B_{\boldsymbol{k}}\right) + \left(A_{\boldsymbol{k}}+B_{\boldsymbol{k}}\right)\left(A^2_{\boldsymbol{k}}-B_{\boldsymbol{k}}^2 -C_{\boldsymbol{k}}^2\right) \right] -i\alpha\omega\left[ \omega^2 + \left(A_{\boldsymbol{k}}+B_{\boldsymbol{k}}\right)^2+C^2_{\boldsymbol{k}} \right],\\
        \end{equation}
\end{subequations}
\end{widetext}
noting that the matrix $\underline{\boldsymbol{W}}^{r,\theta}_{\boldsymbol{k}}$ is Hermitean: $W^{(r,\theta)}_{\beta\alpha,\boldsymbol{k}}(\omega)=W^{\ast(r,\theta)}_{\alpha\beta,\boldsymbol{k}}(\omega)$.

For reference, we include the results for a passive solid composed of elastically linked passive Brownian objects as well.
This system does not exhibit migrating motion. Its constituents do not self-propel.
Since we cannot define an angle of self-propulsion $\theta_i$ in this case, only the fluctuations of two stochastic variables $u_{x,\boldsymbol{k}}$ and $u_{y,\boldsymbol{k}}$ need to be considered.
In this case, the corresponding expressions read
\begin{equation}
    \begin{split}
        S_{xx,\boldsymbol{k}}(\omega)=&\frac{2 D_{r} \left(\omega^2+M_{xy,\boldsymbol{k}}^2+M_{yy,\boldsymbol{k}}^{2}\right)}{\mathrm{(denom.passive)}},\\
        S_{yy,\boldsymbol{k}}(\omega)=&\frac{2 D_{r} \left(\omega^2+M_{xx,\boldsymbol{k}}^2+M_{xy,\boldsymbol{k}}^{2}\right)}{\mathrm{(denom.passive)}},\\
        S_{xy,\boldsymbol{k}}(\omega)=&S_{yx,\boldsymbol{k}}(\omega)\\
        =&- \frac{2 D_r\left(M_{xx,\boldsymbol{k}}+M_{yy,\boldsymbol{k}} \right) M_{xy,\boldsymbol{k}}}{\mathrm{(denom.passive)}},
        \\
        \mathrm{(denom.passive)}=&\omega^4 + \left[M_{xx,\boldsymbol{k}}^{2} + 2 M_{xy,\boldsymbol{k}}^{2} + M_{yy,\boldsymbol{k}}^{2}\right]\omega^2\\
        &+ \left(M_{xx,\boldsymbol{k}} M_{yy,\boldsymbol{k}} - M_{xy,\boldsymbol{k}}^{2}\right)^{2}.
    \end{split}
    \label{eq:passive}
\end{equation}
Both $S_{xx,\boldsymbol{k}}(\omega)$ and $S_{yy,\boldsymbol{k}}(\omega)$ are maximized at $\omega=0$ for any wavevector $\boldsymbol{k}$.
\begin{figure*}
    \centering
    \includegraphics{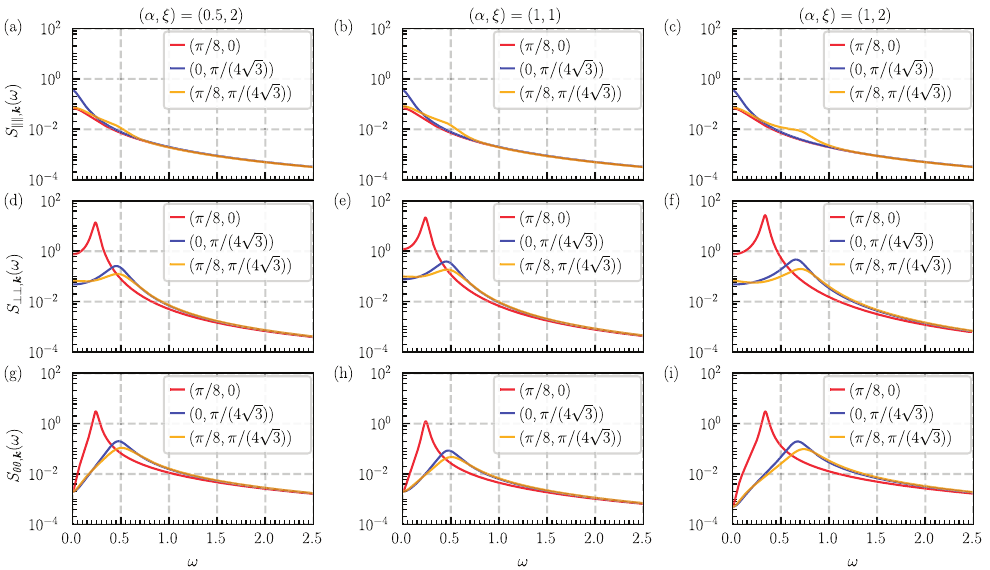}
    \caption{Diagonal components of the matrix of fluctuation spectra $S_{\alpha\alpha,\boldsymbol{k}}(\omega)$ as functions of angular frequency $\omega$ for some fixed wavevectors $\boldsymbol{k}=(k_x,k_y)$ as indicated by the legends and colors of the lines. From the top row, $S_{\parallel\parallel,\boldsymbol{k}}(\omega)$, $S_{\perp\perp,\boldsymbol{k}}(\omega)$, and $S_{\theta\theta,\boldsymbol{k}}(\omega)$ are plotted. Since in our case the solid migrates along the $x$-axis, that is, $\theta_0=0$, we here imply $S_{\parallel\parallel,\boldsymbol{k}}(\omega)=S_{xx,\boldsymbol{k}}(\omega)$ and $S_{\perp\perp,\boldsymbol{k}}(\omega)=S_{yy,\boldsymbol{k}}(\omega)$. 
    (a), (d), and (g) show data for $(\alpha,\xi)=(0.5,2)$; (b), (e), and (h) include data for $(\alpha,\xi)=(1,1)$; and (c), (f), and (i) present data for $(\alpha,\xi)=(1,2)$. }
    \label{fig:S_omega}
\end{figure*}
\begin{figure}
    \centering
    \includegraphics{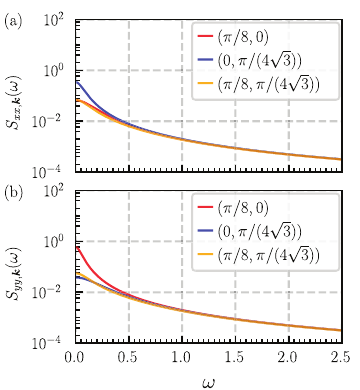}
    \caption{Fluctuation spectra (a) $S_{xx,\boldsymbol{k}}(\omega)$ and (b) $S_{yy,\boldsymbol{k}}(\omega)$ of a corresponding passive solid for comparison. Each colored line includes the data for a given wavevector $\boldsymbol{k}=(k_x,k_y)$ as indicated by the legend.}
    \label{fig:dynamical-structure:passive}
\end{figure}

Hereafter we choose the direction of collective motion as $\theta_0=0$. 
For this configuration, we plot in Fig.~\ref{fig:S_omega} the diagonal components of the matrix of fluctuation spectra $S_{\alpha\alpha,\boldsymbol{k}}(\omega)$, see Eqs.~(\ref{eq:dynamical_structure}), for $\alpha=\parallel,\perp,\theta$ as functions of the angular frequency $\omega$ for some fixed wavevectors $\boldsymbol{k}$.
For comparison, corresponding fluctuation spectra for the situation of a passive solid, see Eqs.~(\ref{eq:passive}), are plotted in Fig.~\ref{fig:dynamical-structure:passive}.
Since $S_{\alpha\alpha,\boldsymbol{k}}(\omega)$ for $\alpha=\parallel,\perp,\theta$ are even functions in $\omega$, we only focus on the region of $\omega\geq 0$.
While the fluctuation spectra of a passive solid show maxima at $\omega=0$ for all considered wavevectors, the ones of an active solid feature maxima at a nonvanishing $\omega$ for $\alpha=\perp,\theta$.
These maxima are thus associated with the activity-induced excitability inside the active solid.

Next, we consider the nonvanishing values of angular frequencies at which the maxima of the fluctuation spectrum are located for $\alpha=\perp,\theta$.  Figure~\ref{fig:S_omega} indicates that these values increase together with the strength of self-propulsion $\alpha$ [compare (a) and (c)] and the strength of alignment $\xi$ [compare (b) and (c)].

Regarding the spatial dependence and anisotropy of the fluctuation spectra, we infer from Figs.~\ref{fig:S_omega} and \ref{fig:dynamical-structure:passive} that both active and passive systems feature larger maximal values for $S_{\parallel\parallel,\boldsymbol{k}}(\omega)$ along the $k_y$-direction than along the $k_x$-direction. For $S_{\perp\perp,\boldsymbol{k}}(\omega)$, it is the other way around. 
Thus, we infer that fluctuations in displacements are enhanced along the directions perpendicular to the wavevector.
$S_{\theta\theta,\boldsymbol{k}}(\omega)$ behaves similarly to $S_{\perp\perp,\boldsymbol{k}}(\omega)$ in terms of the spatial anisotropy.
In our equations of motion, Eqs.~(\ref{eq:dimless:theta}) and (\ref{eq:dimless:omega}), angular deviations are induced by perpendicular displacements $u_{\perp,\boldsymbol{k}}$.

\begin{figure*}
    \centering
    \includegraphics[scale=1]{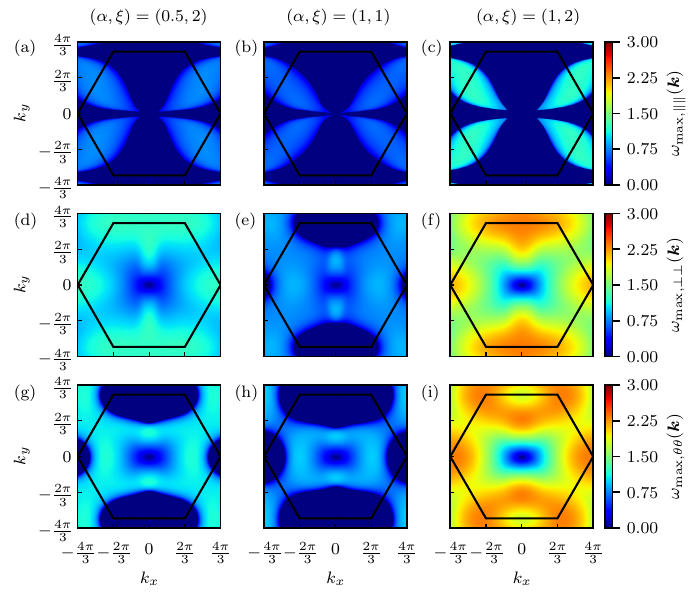}
    \caption{Angular frequencies $\omega_{\mathrm{\max},\alpha\alpha}(\boldsymbol{k})$ that maximize the associated fluctuations $S_{\alpha\alpha,\boldsymbol{k}}(\omega)$ [$\mathrm{max}(S_{\alpha\alpha,\boldsymbol{k}}(\omega))=S_{\alpha\alpha,\boldsymbol{k}}(\omega_{\mathrm{max},\alpha\alpha})$], $\alpha=\parallel,\perp,\theta$, as a function of the wavevector $\boldsymbol{k}$. The strengths of self-propulsion $\alpha$ and of alignment $\xi$ are the same as in Fig.~\ref{fig:S_omega}, that is, from left to right, $(\alpha,\xi)=(0.5,2), (1,1), (1, 2)$. We here set the angle of unperturbed collective migration to $\theta_0=0$.  Dark hexagonal lines indicate the first Brillouin zone.}
    \label{fig:dispersion}
\end{figure*}

\begin{figure*}
    \centering
    \includegraphics[scale=1]{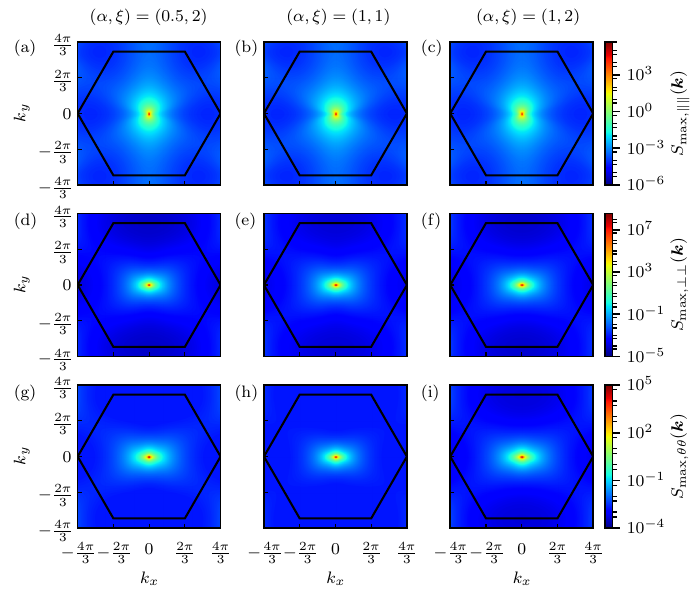}
    \caption{Maximum values of the fluctuation spectra $S_{\mathrm{max},\alpha\alpha}(\boldsymbol{k})$ for $\alpha=\parallel,\perp,\theta$ as a function of the wavevector $\boldsymbol{k}$. The illustration is analogous to the one in Fig.~\ref{fig:dispersion}. Again, dark hexagonal lines indicate the first Brillouin zone.}
    \label{fig:dispersion:max-Sk}
\end{figure*}

For further reference, we denote the angular frequency at the maximum of each diagonal component $S_{\alpha\alpha,\boldsymbol{k}}(\omega)$ as $\omega_{\mathrm{max},\alpha\alpha}(\boldsymbol{k})$ for $\alpha=\parallel,\perp,\theta$. The corresponding maximal value of for $\alpha=\parallel,\perp,\theta$  is obtained via
\begin{equation}
    S_{\mathrm{max},\alpha\alpha}(\boldsymbol{k})\coloneq
    \mathrm{max}\left(S_{\alpha\alpha,\boldsymbol{k}}(\omega)\right)=S\left(\omega_{\mathrm{max},\alpha\alpha}(\boldsymbol{k})\right).
\end{equation}
$\omega_{\mathrm{max},\alpha\alpha}(\boldsymbol{k})$ and $S_{\mathrm{max},\alpha\alpha}(\boldsymbol{k})$ are plotted in Figs.~\ref{fig:dispersion} and \ref{fig:dispersion:max-Sk}, respectively.

According to Fig.~\ref{fig:dispersion}, $\omega_{\mathrm{max},\alpha\alpha}(\boldsymbol{k})$ increases from $\boldsymbol{k}=\boldsymbol{0}$ anisotropically within the $k_x$-$k_y$-plane. In fact, the considered situation is not isotropic because of the set direction of collective migration and the discrete elastic interactions introduced by the springs.
For elevated wavenumbers, we observe in Fig.~\ref{fig:dispersion} maxima of $\omega_{\mathrm{max},\theta\theta}(\boldsymbol{k})$ along each direction of $\boldsymbol{k}$. 
The positions of these maxima shift toward higher wavenumbers as $\alpha$ [compare (a) and (c)] or $\xi$ [compare (b) and (c)] increase.
Interestingly, therefore, fluctuations of finite nonvanishing angular frequency show maxima in the considered overdamped systems where classical phonons do not appear by construction.
Thus, in contrast to the corresponding passive case, a significant fraction of fluctuations propagates through the system with characteristic nonvanishing angular frequencies.

Finally, we focus on the dependence of the maxima of the fluctuation spectra on the wavevector $\boldsymbol{k}$, see Fig.~\ref{fig:dispersion:max-Sk}.
The magnitudes in the spectra show maxima at zero wavenumber and then rapidly decay, suggesting that the fluctuations in this system mainly feature long-wavelength components.
In both active and passive systems, $\omega_{\mathrm{max},\alpha\alpha}(\boldsymbol{k}=\boldsymbol{0})=0$ is due to the presence of Goldstone modes, see Fig.~\ref{fig:dispersion}. 
In fact, the mode $\boldsymbol{k}=\boldsymbol{0}$ does not imply any elastic deformation of the springs and thus there is no restoring force.

\section{\label{sec:entropy} Rate of entropy production}
We characterize the collective excitation with respect to the energy dissipation as in Ref.~\onlinecite{Caprini2023-ro}.
The rate of entropy production~\cite{Spinney2012-rk,Dabelow2019-zi,Seifert2012-ri} is defined as
\begin{equation}
    \dot{s}=\lim_{T\rightarrow \infty}\frac{1}{T}\left\langle \ln\frac{\mathcal{P}(T)}{\mathcal{P}_{r}(T)}\right\rangle,
\end{equation}
where $\braket{\cdots}$ is the ensemble average. $P$ and $P_r$ are the forward and backward trajectories scaling as
\begin{equation}
    \begin{split}
        \mathcal{P}(T)\sim&\;\mathrm{e}^{-\mathcal{A}(T)},\\
        \mathcal{P}_r(T)\sim&\;\mathrm{e}^{-\mathcal{A}_{r}(T)},
    \end{split}
\end{equation}
where $\mathcal{A}$ and $\mathcal{A}_r$ represent the forward and backward actions, respectively. They are calculated via
\begin{equation}
\begin{split}
    \mathcal{A}(T)=&\frac{1}{2}\sum_{\boldsymbol{k}}\int_{-\infty}^{\infty} dt\, W_T(t)\left|\boldsymbol{L}_{\boldsymbol{k}}(t)\right|^2,\\
     \mathcal{A}_r(T)=&\frac{1}{2}\sum_{\boldsymbol{k}}\int_{-\infty}^{\infty} dt\, W_T(t)\left|\boldsymbol{L}_{r,\boldsymbol{k}}(t)\right|^2.
\end{split}
\end{equation}
The summation over $\boldsymbol{k}$ implies that all modes contribute to the total entropy production. $W_T(t)$ is a rectangular window function defined by 
\begin{equation}
    W_T(t)=
    \left\{
    \begin{array}{cc}
         1 & \mathrm{if }-T\leq t \leq T, \\
         0 &  \mathrm{otherwise},
    \end{array}
    \right.
\end{equation}
and $\boldsymbol{L}_{(r),\boldsymbol{k}}(t)$ represent the inverse Fourier transformations of $\boldsymbol{L}_{(r),\boldsymbol{k}}(\omega;T)$, obtained by solving for the stochastic contributions in the formal solutions of the equation of motion, Eq.~(\ref{eq_equationofmotion}), via
\begin{equation}
    \boldsymbol{L}_{\boldsymbol{k}}(\omega;T)={\underline{\boldsymbol{D}}^{(\theta_0)}}^{-1}\cdot\left[ -i\omega\underline{\boldsymbol{I}}+\underline{\boldsymbol{\Gamma}}^{(\theta_0)}_{\boldsymbol{k}} \right]\cdot\tilde{\boldsymbol{z}}_{\boldsymbol{k}}(\omega;T),
\end{equation}
\begin{align}
    \boldsymbol{L}_{r,\boldsymbol{k}}(\omega;T)&=\mathcal{T}\boldsymbol{L}_{\boldsymbol{k}}(\omega;T)\nonumber\\
    &={\underline{\boldsymbol{D}}^{(\theta_0)}}^{-1}\cdot\left[ i\omega\underline{\boldsymbol{I}}+\underline{\boldsymbol{\Gamma}}^{(\theta_0)}_{\boldsymbol{k}} \right]\cdot\tilde{\boldsymbol{z}}_{\boldsymbol{k}}(\omega;T).
\end{align}
In the latter equation, $\mathcal{T}$ is the time-reversal operator.
We here consider the dynamics within the finite time window $[-T,T]$ as indicated by the additional parameter $T$ in the arguments of the functions. 

Combining the above expressions and applying the Wiener--Khinchin theorem, see Appendix~\ref{appendix:wiener-khinchin}, we find the spectral decomposition of the entropy production $\sigma(\omega, \boldsymbol{k})$, defined via
\begin{equation}
    \dot{s}=\int_{-\infty}^{\infty}\frac{d\omega}{2\pi}\sum_{\boldsymbol{k}}\sigma_{\boldsymbol{k}}(\omega),
    \label{eq:dots}
\end{equation}
as
\begin{align}
    \sigma_{\boldsymbol{k}}(\omega)=&\lim_{T\rightarrow\infty}\frac{1}{2T}\left\langle \left|\boldsymbol{L}_{r,\boldsymbol{k}}(\omega;T)\right|^2-\left|\boldsymbol{L}_{\boldsymbol{k}}(\omega;T)\right|^2 \right\rangle \nonumber\\    =&\lim_{T\rightarrow\infty}\frac{1}{2T}2i\omega\left\langle\tilde{\boldsymbol{z}}^T_{\boldsymbol{k}}(\omega;T)\cdot \left[\left(\underline{\boldsymbol{B}}\cdot\underline{\boldsymbol{\Gamma}}^{(s)}_{\boldsymbol{k}}-\underline{\boldsymbol{\Gamma}}^{(s)}_{\boldsymbol{k}}\cdot\underline{\boldsymbol{B}}\right)\right.\right.\nonumber\\
&\left.\left.+\left(\underline{\boldsymbol{B}}\cdot\underline{\boldsymbol{\Gamma}}^{(a)}_{\boldsymbol{k}}+\underline{\boldsymbol{\Gamma}}^{(a)}_{\boldsymbol{k}}\cdot\underline{\boldsymbol{B}}\right)\right]\cdot\tilde{\boldsymbol{z}}_{-\boldsymbol{k}}(-\omega;T)\right\rangle\nonumber\\
    =&-2\pi N\frac{2\omega\alpha\left( \xi^2D_r+D_{\theta} \right)}{D_r D_{\theta}}\mathrm{Im}\left[S_{\theta\perp,\boldsymbol{k}}(\omega)\right]\nonumber\\
=&2\pi N\frac{4\alpha^2\omega^2\left(\omega^2+\left(A_{\boldsymbol{k}}+B_{\boldsymbol{k}}\right)^2+C_{\boldsymbol{k}}^2\right)}{\mathrm{(denom.)}}
    \nonumber\\&\qquad\times    \frac{\left(\xi^2D_r+D_{\theta}\right)^2}{D_rD_{\theta}}.
    \label{eq:spectral-entropy}
\end{align}
We here used Eq.~\eqref{eq:dynamical_structure}. Moreover, we denote by $\underline{\boldsymbol{\Gamma}}^{(s)}_{\boldsymbol{k}}$ and $\underline{\boldsymbol{\Gamma}}^{(a)}_{\boldsymbol{k}}$ the symmetric and antisymmetric parts of the matrix $\underline{\boldsymbol{\Gamma}}^{(\theta_0)}_{\boldsymbol{k}}$, respectively, defined via
\begin{align}
    \underline{\boldsymbol{\Gamma}}^{(s)}_{\boldsymbol{k}}=&\frac{1}{2}\left(\underline{\boldsymbol{\Gamma}}^{(\theta_0)}_{\boldsymbol{k}}+{\underline{\boldsymbol{\Gamma}}^{(\theta_0)}_{\boldsymbol{k}}}^T\right),\\
    \underline{\boldsymbol{\Gamma}}^{(a)}_{\boldsymbol{k}}=&\frac{1}{2}\left(\underline{\boldsymbol{\Gamma}}^{(\theta_0)}_{\boldsymbol{k}}-{\underline{\boldsymbol{\Gamma}}^{(\theta_0)}_{\boldsymbol{k}}}^T\right).
\end{align}
Besides, we introduced the abbreviation
\begin{equation}
    \begin{split}
\underline{\boldsymbol{B}}=&\left({{\underline{\boldsymbol{D}}^{(\theta_0)}}^{-1}}\right)^{T}\cdot{\underline{\boldsymbol{D}}^{(\theta_0)}}^{-1}\\
    =&\begin{pmatrix}
        \frac{1}{2D_r} &  0 &  0\\
        0 &  \frac{1}{2D_r}+\frac{\xi^2}{2D_\theta} &  -\frac{\xi}{2D_\theta}\\
        0 &  -\frac{\xi}{2D_\theta} &  \frac{1}{2D_\theta}\\
    \end{pmatrix}
    \end{split}
\end{equation}

\begin{figure}
    \centering
    \includegraphics{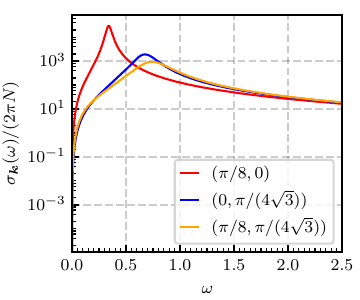}
    \caption{Spectral entropy production $\sigma_{\boldsymbol{k}}(\omega)$ for fixed wavevectors $\boldsymbol{k}=(k_x,k_y)$ as indicated by the color of the lines and in the legend. Again, the direction of collective migration is set along the $x$-direction, that is, $\theta_0=0$.}
    \label{fig:entropy-prod:1d}
\end{figure}

We plot in Fig.~\ref{fig:entropy-prod:1d} the spectral entropy production $\sigma_{\boldsymbol{k}}(\omega)$ as a function of angular frequency $\omega$ for different wavevectors $\boldsymbol{k}$. Obviously, $\sigma_{\boldsymbol{k}}(\omega)$ features maxima at nonvanishing angular frequency $\omega$ similarly to the fluctuation spectra, see Fig.~\ref{fig:S_omega}. 
We denote the angular frequency of the maximum for a given wavevector $\boldsymbol{k}$ as $\omega_{\mathrm{max},\sigma}(\boldsymbol{k})$.
For the considered wavevectors $\boldsymbol{k}$, the positions of $\omega_{\mathrm{max},\sigma}(\boldsymbol{k})$ are closer to the ones of $\omega_{\mathrm{max},\perp\perp}(\boldsymbol{k})$ and $\omega_{\mathrm{max},\theta\theta}(\boldsymbol{k})$ than $\omega_{\mathrm{max},\parallel\parallel}(\boldsymbol{k})$, see Fig.~\ref{fig:S_omega}. 
These relations are conceivable because the coupling between $u_\perp$ and $\delta\theta$ significantly contributes to the spectral entropy production. The coupling becomes apparent from the emergence of $S_{\theta\perp,\boldsymbol{k}}(\omega)$ in Eq.~(\ref{eq:spectral-entropy}).

\begin{figure}
    \centering
    \includegraphics{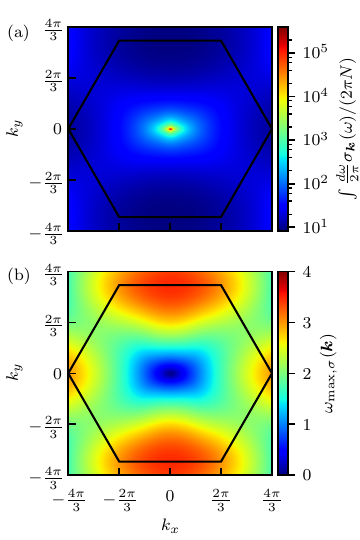}
    \caption{(a) Spectral entropy production $\sigma_{\boldsymbol{k}}(\omega)$ integrated over angular frequency $\omega$ to yield the contributions to the rate of entropy production $\dot{s}$, see Eq.~(\ref{eq:dots}). (b) Angular frequency $\omega_{\mathrm{max},\boldsymbol{k}}(\omega)$ that maximizes $\sigma_{\boldsymbol{k}}(\omega)$. Both are displayed in the $k_x$-$k_y$-plane of the components of the wavevector $\boldsymbol{k}$.  
    Dark hexagonal lines indicate the first Brillouin zone.}
    \label{fig:entropy-prod:2d}
\end{figure}

The spectral entropy production $\sigma_{\boldsymbol{k}}(\omega)$ is integrated over $\omega$, $\int \sigma_{\boldsymbol{k}}(\omega)\,{d\omega}/{2\pi}$, to obtain the contributions per wavevector $\boldsymbol{k}$ to the rate of entropy production $\dot{s}$, see Eq.~(\ref{eq:dots}). It is plotted in Fig.~\ref{fig:entropy-prod:2d}(a), while $\omega_{\mathrm{max},\sigma}(\boldsymbol{k})$ is included in Fig.~\ref{fig:entropy-prod:2d}(b).
We infer from Fig.~\ref{fig:entropy-prod:2d}(a) that the integrated spectral entropy production sharply decreases as the wavenumber $|\boldsymbol{k}|$ increases.
Moreover, it decays faster in $k_y$-direction than in $k_x$-direction.
Besides, Fig.~\ref{fig:entropy-prod:2d}(b) shows that fluctuations of larger angular frequency $\omega$ tend to contribute more significantly to the entropy production with increasing $|\boldsymbol{k}|$.

\section{Numerical evaluation \label{sec:numeric}}
At the beginning of the calculation of the fluctuation spectrum in Sec.~\ref{sec:structure}, we assumed that the fluctuations are small enough to linearize the equations and that $V=\alpha$.
To check whether these assumptions are correct, we numerically implemented the original equations, Eqs.~\eqref{eq:dimless:r}--\eqref{eq:dimless:omega}, in rectangular calculation boxes.
The number of self-propelled objects is set to $N=256\times 256$. 
We employ a Cartesian coordinate system and denote the dimensions of the system by box lengths $L_x=256$ and $L_y=256\times\sqrt{3}/2$, see Fig.~\ref{fig:system-schematic}. 
Periodic boundary conditions are imposed along all edges of the rectangular calculation box. 
The self-migrating objects are initially arranged on regular spring-lattice structures of perfect hexagonal symmetry.
Moreover, initially, the directions $\boldsymbol{e}_i$ ($i=1,...,N$) of self-propulsion head towards the $+\hat{\boldsymbol{\mathrm{x}}}$-direction.  
The equations of motion, Eqs.~\eqref{eq:dimless:r}--\eqref{eq:dimless:omega} are numerically integrated forward in time using the Euler--Maruyama method. We use a time step of $dt=0.001$ and the parameter values in the rescaled Eqs.~\eqref{eq:dimless:r}--\eqref{eq:dimless:omega} as listed in Table~\ref{tab:parameters}. 

Time evolutions of appropriate parameters are employed to check when the system has reached a steady state. 
For later discussion, we define the current polar order parameter
\begin{equation}
\label{eq:polarorder}
    \boldsymbol{p}(t)=\frac{1}{N}\sum_{i=1}^N\boldsymbol{e}(\theta_{i}(t))
\end{equation}
and the current velocity of the center of mass
\begin{equation}
\label{eq:speed_of_solid}
    \boldsymbol{v}_c(t)=\frac{1}{N}\sum_{i=1}^N\boldsymbol{v}_i(t).
\end{equation}

\begin{figure}
    \centering
    \includegraphics{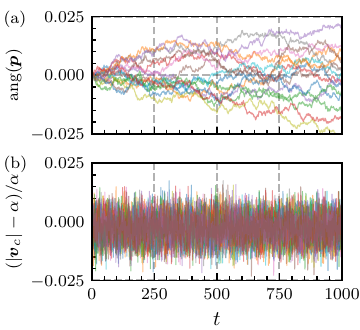}
    \caption{Time series of (a): the angle of temporal polar order parameter and (b): the speed of the center of the mass. There are 16 lines obtained from different seeds for stochastic forces.}
    \label{fig:numerical:time-series}
\end{figure}

The angular orientation of $\boldsymbol{p}(t)$, denoted as $\mathrm{ang}(\boldsymbol{p}(t))$, and the relative deviation of the magnitude of $\boldsymbol{v}_c$ from the parameter $\alpha$, that is, $(|\boldsymbol{v}_c(t)|-\alpha)/\alpha$, see Eqs.~(\ref{eq:polar-order}) and (\ref{eq:V-alpha}), are plotted in Fig.~\ref{fig:numerical:time-series} as a function of time for $16$ different numerical realizations.
Since the numerical system is finite-sized, the fluctuations of $\mathrm{ang}(\boldsymbol{p})$ remain finite (nonzero), yet within a small range of $[-0.025,0.025]$ within the considered time window, see Fig.~\ref{fig:numerical:time-series}(a).
Moreover, in our simulations, the maximum fluctuations of the relative speeds of the centers of the mass $(|\boldsymbol{v}_c(t)|-\alpha)/\alpha$ stay within $2.5\;$\% for all samples within the considered time window, see Fig.~\ref{fig:numerical:time-series}(b).
Thus, we may conclude that they remain small enough for our approximations to be valid. 

Statistically, the degree of overall orientational order and the speed of the center of the mass are then calculated as
\begin{equation}
    \begin{split}
        \boldsymbol{P}=&\langle \boldsymbol{p} \rangle_t,\\
    \end{split}
\end{equation}
and
\begin{equation}
    \begin{split}
        \boldsymbol{V}=&\langle \boldsymbol{v}_c \rangle_t,\\
    \end{split}
\end{equation}
respectively, where we use the subscript $_t$ to mark numerical ensemble and time average.
These averages are taken over the 16 considered numerical realizations iterated from different seeds for the stochastic forces and over the time window $(500, 1000]$, when the system has reached a steady state. Within this time window, data are used for averaging from all integer times $t=501, 502, ..., 1000\,$.
The resulting values are $|\boldsymbol{P}|=0.997315\pm 0.000254$ and $(|\boldsymbol{V}|-\alpha)/\alpha=(-2.647\pm5.574)\times 10^{-3}$. 
\begin{table}[b]
\caption{\label{tab:parameters}
Values of the rescaled parameters set in our numerical simulation.
}
\begin{ruledtabular}
\begin{tabular}{ccc}
\textrm{Parameter}&
\textrm{Description}&
\textrm{Value(s)}\\
\colrule
$N$ & number of objects & $256\times 256$\\
$\alpha$ & strength of self-propulsion & $1$\\
$\xi$ & strength of self-alignment & $2$\\
$D_r$ & translational diffusion coefficient & $0.001$\\
$D_\theta$& rotational diffusion coefficient & $0.001$
\end{tabular}
\end{ruledtabular}
\end{table}
Besides, we numerically confirmed that the distribution of $\delta\theta_i(t)=\theta_i(t)-\mathrm{ang}(\boldsymbol{p}(t))$ is steady after $t=500$.
It fits well with a Gaussian curve of standard deviation $7.5\times 10^{-2}$, indicating on average only moderate angular deviations of the individual migrating objects from the common migration direction.

Numerically, $\boldsymbol{z}_{i}(t)$ for $i=1,..., N$ are obtained in the following way. 
We determine the reference positions that correspond to an undeformed state, in which the induced deformations of the mesh of springs vanish. From there, the positional displacements are determined. Moreover, the angular deviations are measured as deviations from the direction of $\mathrm{ang}(\boldsymbol{p}(t))$. These evaluations are performed for each instance of time $t$ of numerical measurement. 

\begin{figure}
    \centering
    \includegraphics{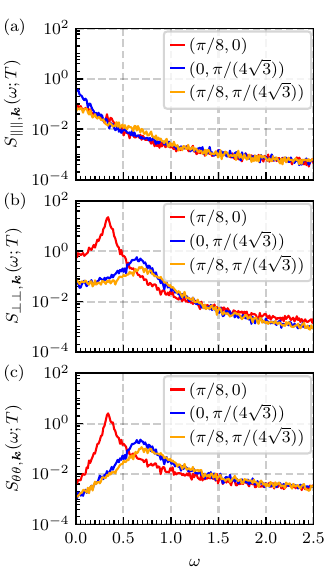}
    \caption{Numerically calculated fluctuation spectra $S_{\beta\beta,\boldsymbol{k}}(\omega;T)$, see Eq.~(\ref{eq:S_numerics}), where $\beta\in\{\parallel,\perp,\theta\}$, for strengths of self-propulsion and angular alignment $(\alpha,\xi)=(1,2)$. Results are obatained from numerical measurements on $N=256\times256$ objects, see Table~\ref{tab:parameters}, sampling over $16$ numerical realizations and a time window of width $2T=500$. Different wavevectors $\boldsymbol{k}=(k_x,k_y)$ are selected as indicated in the legends and by the different line colors. The resulting curves match well in shape the analytically calculated ones in Fig.~\ref{fig:S_omega}(c,f,i).}
    \label{fig:numerical:fluctuation}
\end{figure}

From the prescribed time window of width $2T$, the resulting functions $S_{\alpha\alpha,\boldsymbol{k}}(\omega;T)$ corresponding to the fluctuation spectra are determined, where $\alpha\in\{\parallel,\perp,\theta\}$.
To this end, we first transform the discrete time series of numerical measurement data ${z}_{\alpha,i}(t)$ confined to the time window of width $2T$ to Fourier space, yielding $\tilde{z}_{\alpha, \boldsymbol{k}}(\omega;T)$. We list the general definitions of functions of arguments $(\omega;T)$ and of the discrete Fourier series in Appendix~\ref{appendix:fourier}. 
The expressions $S_{\alpha\alpha,\boldsymbol{k}}(\omega;T)$ are found from ensemble averages $\braket{...}_{\mathrm{ens}}$ as
\begin{equation}
    S_{\alpha\alpha, \boldsymbol{k}}(\omega;T)=\frac{\left\langle \tilde{z}_{\alpha,\boldsymbol{k}}(\omega;T)\tilde{z}_{\alpha,\boldsymbol{-k}}(-\omega;T) \right\rangle_{\mathrm{ens}}}{2TN}.
    \label{eq:S_numerics}
\end{equation}
Corresponding results are plotted in Fig.~\ref{fig:numerical:fluctuation}
and match well in shape with the analytical ones, see Fig.~\ref{fig:S_omega}.

\section{Conclusions \label{sec:conclusion}}
In summary, we analyze the excitability of the orientationally ordered, collectively moving state of active solids consisting of self-propelling objects that are elastically linked to each other.
To this end, the fluctuation spectra for the elastic displacements $\boldsymbol{u}$ and deviations in the angle of self-propulsion $\theta$ are calculated from the underlying equations.

The self-correlations, apparent in the diagonal components of the matrix representing the fluctuation spectra, contain the spatiotemporal information on these excitations.
In the case of passive solids, they decay monotonically as functions of frequency.
In contrast to that, the fluctuation spectra for collectively migrating active solids feature maxima at a nonvanishing frequency and a nonvanishing wavenumber. They seem to be related to associated modes of entropy production.

In contrast to that, we find that entropy production is related to the off-diagonal entries of the matrix representing the fluctuation spectra.
It was stated before that entropy production in steady states stems from the nonreciprocal components of dynamical equations~\cite{Loos2020-jt}.
The rate of entropy production for the considered orientationally ordered, collectively migrating active solids is maximized at vanishing wavenumber. However, for fixed nonvanishing wavenumbers, the rate of entropy production is maximized at nonvanishing frequency. 

In our work, we address additional coupling between elastic forces on individual self-propelling objects and their direction of active migration. Previous studies already considered excitations of active solids, yet in the absence of such migrational alignment~\cite{Caprini2019-oe}. 
We focus on excitations in the collectively moving, orientationally ordered state, while previous work addressed static correlation functions~\cite{Huang2021-js}.

Overall, we hope to stimulate by our study corresponding experimental realizations and observations. A challenge is certainly to produce large enough systems for quantitative matching, particularly in terms of associated Hexbug$^\text{\textregistered}$ experiments \cite{Baconnier2022-lf,Baconnier2023-qg,Hernandez-Lopez2024-mx}, although we are confident that basic features can be reproduced qualitatively by focusing on the inner parts of larger collections. If direct comparison is not possible due to a dominating role of boundaries, one strategy may consist in employing programmable self-propelling robots \cite{Rubenstein2014-ho}. By tracking their migration and controlling them from outside to mimic and apply mutual elastic interactions between them and adjusting their motion accordingly, periodic boundary conditions could be imposed at the same time.

\begin{acknowledgments}
Y.K.\ acknowledges support of the work through the International Joint Graduate Program in Materials Science at Tohoku University and JST SPRING, Grant Number JPMJSP2114. A.M.M.\ thanks the Deutsche Forschungsgemeinschaft (German Research Foundation, DFG) for the Heisenberg Grant no.\ ME 3571/4-1. Moreover, support by the Deutsche Forschungsgemeinschaft through the Research Grant no.\ ME 3571/12-1 is gratefully acknowledged. 
\end{acknowledgments}

\section*{Author Declarations}
\subsection*{Conflict of Interest}
The authors have no conflicts to disclose.

\subsection*{Author Contributions}
Y.K.\ and A.M.M.\ designed the research. Y.K.\ performed the initial analytical calculations. N.U.\ and A.M.M.\ checked the analytical calculations. Y.K.\ performed the numerical simulations. All authors discussed the results. Y.K.\ wrote the initial draft of the manuscript. A.M.M.\ iterated the manuscript. All authors were involved in finalizing the manuscript.

\section*{Data Availability}
The data supporting the results in this study are available from the corresponding author upon reasonable request. Results obtained from the analytical expressions are plotted directly. Data output from the numerical simulations is displayed in the corresponding figures. 

\appendix

\section{Fourier transformation \label{appendix:fourier}}

In this appendix, we list the conventions that we use to perform our Fourier transformations. 

\subsection{Fourier transformation in time}
The Fourier transformation of the function $\boldsymbol{f}=\boldsymbol{f}(t)$ in time is defined in the form
\begin{subequations}
    \begin{equation}
    \tilde{\boldsymbol{f}}(\omega)\coloneqq\mathcal{F}\{\boldsymbol{f}\}(\omega)=\int_{-\infty}^{\infty} dt \boldsymbol{f}(t)\mathrm{e}^{i\omega t},
  \end{equation}
and its inverse via 
  \begin{equation}
    \boldsymbol{f}(t)=\mathcal{F}^{-1}\{\tilde{\boldsymbol{f}}\}(t)=\frac{1}{2\pi}\int_{-\infty}^{\infty} d\omega \tilde{\boldsymbol{f}}(\omega)\mathrm{e}^{-i\omega t}.
  \end{equation}
\end{subequations}
For our purpose, we further define the Fourier transformation of $\boldsymbol{f}(t)$ confined to the time window $[-T,T]$ as 
\begin{equation}
    \begin{split}
        \tilde{\boldsymbol{f}}(\omega;T)\coloneqq \mathcal{F}\{W_T(t)\boldsymbol{f}\}(\omega) =\int_{-\infty}^{\infty} dt \,W_T(t)\boldsymbol{f}(t)\mathrm{e}^{i\omega t},
    \end{split}
\end{equation}
where $W_T(t)$ is given by 
\begin{equation}
    W_T(t)=
    \left\{
    \begin{array}{cl}
         1 & \mathrm{ if } -T\leq t \leq T, \\[.2cm]
         0 &  \mathrm{otherwise}.
    \end{array}
    \right.
\end{equation}

In the discrete case of a time series of measurement data $\left\{a_m\right\}=a_0, ..., a_{n-1}$, sampled at $n$ equally distanced instances in time as indexed by $m\in\{0,1,...,n-1\}$, the Fourier series is written in the form
\begin{subequations}
\begin{equation}
    \begin{split}
        \tilde{a}_k \coloneq \sum_{m=0}^{n-1}a_m\mathrm{e}^{2\pi i\frac{mk}{n}} 
    \end{split}
\end{equation}
for $k=0,..., n-1$. 
Likewise, its inverse results via
\begin{equation}
    \begin{split}
        a_m = \frac{1}{n}\sum_{k=0}^{n-1}\tilde{a}_k\mathrm{e}^{-2\pi i\frac{mk}{n}}
    \end{split}
\end{equation}
for $m=0,..., n-1$. 
\end{subequations}

\subsection{Fourier series in space}

When introducing the discrete Fourier series, two possibilities come into our mind that both have their justification. First, when assuming periodic boundary conditions in a rectangular calculation box, we may use the discrete wavevectors defined by standing waves within the calculation box. Second, we may work with the discrete lattice vectors determined by the hexagonal lattice in its undeformed state. In the following, we adopt the latter point of view. 

The arbitrary lattice points are expressed in terms of a linear combination of two unit lattice vectors,
\begin{equation}
        \boldsymbol{R}= m_1\boldsymbol{a}_1+m_2\boldsymbol{a}_2~,
    \end{equation}
 where $m_1=1, ..., N_1$ and $m_2=1,..., N_2$. We choose 
    \begin{equation}
      \boldsymbol{a}_1 = \left(1, 0\right)^T,~      \boldsymbol{a}_2 = \left(\frac{1}{2}, \frac{\sqrt{3}}{2}\right)^T.
      \nonumber
    \end{equation}
Our inverse unit lattice vectors are defined as
\begin{align}
    \boldsymbol{b}_1=&2\pi\frac{\boldsymbol{a}_2\times\boldsymbol{a}_3}{\boldsymbol{a}_1\cdot\left(\boldsymbol{a}_2\times\boldsymbol{a}_3\right)}=2\pi\left(1,-\frac{1}{\sqrt{3}}\right)^T,\\
   \boldsymbol{b}_2=&2\pi\frac{\boldsymbol{a}_3\times\boldsymbol{a}_1}{\boldsymbol{a}_2\cdot\left(\boldsymbol{a}_3\times\boldsymbol{a}_1\right)}=2\pi\left(0,\frac{2}{\sqrt{3}}\right)^T,
\end{align}
where for the ease of derivation we considered an additional third dimension introducing $\boldsymbol{a}_3=(0,0,1)^T$.

The arbitrary inverse lattice points are expressed via
\begin{equation}
    \boldsymbol{G}=\frac{n_1}{N_1}\boldsymbol{b}_1+\frac{n_2}{N_2}\boldsymbol{b}_2,
\end{equation}
$n_1=1,...,N_1$ and $n_2=1,..., N_2$.
Then, the discrete Fourier transformation of $f(m_1,m_2)$ is defined as
\begin{equation}
\begin{split}
    \tilde{f}(n_1,n_2)\coloneq&\mathcal{F}\left[f(m_1,m_2)\right]\\
    =&\sum_{m_1=1}^{N_1} \sum_{m_2=1}^{N_2} \mathrm{e}^{-i\boldsymbol{R}\cdot\boldsymbol{G}}f(m_1,m_2)\\
    =&\sum_{m_1=1}^{N_1} \sum_{m_2=1}^{N_2} \mathrm{e}^{-i\left(\frac{m_1n_1}{N_1}\boldsymbol{a}_1\cdot\boldsymbol{b_1} + \frac{m_2n_2}{N_2}\boldsymbol{a}_2\cdot\boldsymbol{b_2} \right)}
    \\[-.3cm]
    &\qquad\qquad\qquad\qquad\times f(m_1,m_2)\\
    =&\sum_{m_1=1}^{N_1} \sum_{m_2=1}^{N_2} \mathrm{e}^{-2\pi i\left(\frac{m_1n_1}{N_1} + \frac{m_2n_2}{N_2} \right)}f(m_1,m_2),
\end{split}
\end{equation}
and its inverse as
\begin{equation}
\begin{split}
    f(m_1,m_2)=&\mathcal{F}^{-1}\left[\tilde{f}(n_1,n_2)\right]\\
    =&\frac{\sum_{n_1=1}^{N_1} \sum_{n_2=1}^{N_2}}{N_1N_2} \mathrm{e}^{i\boldsymbol{R}\cdot\boldsymbol{G}}\tilde{f}(n_1,n_2)\\
    =&\frac{\sum_{n_1=1}^{N_1} \sum_{n_2=1}^{N_2}}{N_1N_2} \mathrm{e}^{i\left(\frac{m_1n_1}{N_1}\boldsymbol{a}_1\cdot\boldsymbol{b_1} + \frac{m_2n_2}{N_2}\boldsymbol{a}_2\cdot\boldsymbol{b_2} \right)}\\[-.3cm]
    &\qquad\qquad\qquad\qquad\times \tilde{f}(n_1,n_2)\\
    =&\frac{\sum_{n_1=1}^{N_1} \sum_{n_2=1}^{N_2}}{N_1N_2} \mathrm{e}^{2\pi i\left(\frac{m_1n_1}{N_1} + \frac{m_2n_2}{N_2} \right)}\tilde{f}(n_1,n_2).
\end{split}
\end{equation}
A convenient relation is given by
\begin{equation}
\begin{split}
    \frac{\sum_{m_1=1}^{N_1}\sum_{m_2=1}^{N_2}}{N_1N_2}&\mathrm{e}^{-2\pi i\left(\frac{m_1(n_1-n_1^\prime)}{N_1}+\frac{m_2(n_2-n_2^\prime)}{N_2}\right)}\\
    &=\delta_{n_1,n_1^\prime}\delta_{n_2,n_2^\prime},
\end{split}
\end{equation}
where $\delta$ represents the Kronecker delta. 

In a slight variation of notation, we express the Fourier series, here for the displacements $\boldsymbol{u}_i$ ($i=1, ..., N$), as
\begin{equation}
    \boldsymbol{u}_{\boldsymbol{k}}=\sum_{i=1}^N\mathrm{e}^{-\mathrm{i}\boldsymbol{k}\cdot\boldsymbol{x}^{(0)}_i}\boldsymbol{u}_i,
\end{equation}
where $\boldsymbol{x}_i^{(0)}$ is the position vector of the $i$~th object in the undeformed state of the lattice.
Its inverse follows as
\begin{equation}
\boldsymbol{u}_{i}=\frac{1}{N}\sum_{\boldsymbol{k}}\mathrm{e}^{\mathrm{i}\boldsymbol{k}\cdot\boldsymbol{x}^{(0)}_i}\boldsymbol{u}_{\boldsymbol{k}},
\end{equation}
where the summation runs over the wavevectors $\boldsymbol{k}$.

A convenient relation is given by
\begin{equation}
    \delta_{\boldsymbol{k},\boldsymbol{k}^\prime}=\frac{1}{N}\sum_{i=1}^N\mathrm{e}^{\mathrm
-{i}\left( \boldsymbol{k}-\boldsymbol{k}^\prime \right)\cdot\boldsymbol{x}_i^{(0)}},
\end{equation}
where $\delta$ represents the Kronecker delta. 

\section{Definition and derivation of the elasticity matrix in Fourier space \label{appendix:elastic}}

\begin{figure}
    \centering
    \includegraphics{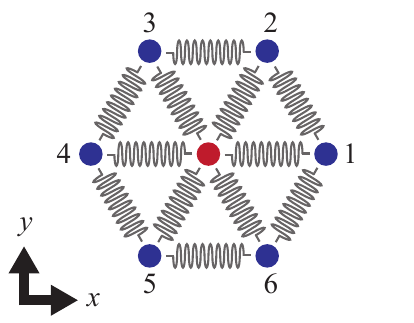}
    \caption{Hexagonal lattice structure around one centered object (red) in the undeformed ground state of the lattice, together with the orientation of our two-dimensional Cartesian coordinate system. There are six nearest-neighboring objects (blue) surrounding the centered object. The situation is identical for each object in the undeformed hexagonal structure.}
    \label{fig:hexagonal}
\end{figure}

The elastic force acting on the object $i$ is 
\begin{equation}
    \boldsymbol{F}_i=-\sum_{j\in (i,j)}\underline{\boldsymbol{M}}_{ij}\cdot\boldsymbol{u}_{ij},
    \label{eq:Fi:appB}
\end{equation}
where
\begin{equation}
    \underline{\boldsymbol{M}}_{ij}=
    \begin{pmatrix}
        \left(x_{ij}^{(0)}\right)^2 & x_{ij}^{(0)} y_{ij}^{(0)}\\[.1cm]
        x_{ij}^{(0)} y_{ij}^{(0)} & \left(y_{ij}^{(0)}\right)^2.
    \end{pmatrix},
\end{equation}
see Eqs.~\eqref{eq:elasticforce:realspace} and~\eqref{eq:elastic:realspace}. 
Multiplying Eq.~\eqref{eq:Fi:appB} by the factor $\mathrm{e}^{-i\boldsymbol{k}\cdot\boldsymbol{x}^{(0)}_i}$ and summing over $i=1,..., N$, we obtain the Fourier transformed expression of $\boldsymbol{F}_i$, 
\begin{equation}
    \begin{split}
        \boldsymbol{F}_{\boldsymbol{k}}=&-\sum_{i=1}^N\sum_{j\in (i,j)}\mathrm{e}^{-i\boldsymbol{k}\cdot\boldsymbol{x}^{(0)}_i}\underline{\boldsymbol{M}}_{ij}\cdot\boldsymbol{u}_{ij}\\
        =&-\sum_{i=1}^N\sum_{j\in (i,j)}\mathrm{e}^{-i\boldsymbol{k}\cdot\boldsymbol{x}^{(0)}_i}\underline{\boldsymbol{M}}_{ij}\cdot\left[ \frac{1}{N}\sum_{\boldsymbol{k}_1}\mathrm{e}^{i\boldsymbol{k}_1\cdot\boldsymbol{x}^{(0)}_i}\boldsymbol{u}_{\boldsymbol{k}_1} \right.
        \\
        & \qquad \left. - \frac{1}{N}\sum_{\boldsymbol{k}_2}\mathrm{e}^{i\boldsymbol{k}_2\cdot\boldsymbol{x}^{(0)}_j}\boldsymbol{u}_{\boldsymbol{k}_2}\right]\\
        =&-\sum_{\boldsymbol{k}_1}\sum_{i=1}^N\sum_{j\in (i,j)}\underline{\boldsymbol{M}}_{ij}\cdot\frac{1}{N}\mathrm{e}^{-i\left(\boldsymbol{k}-\boldsymbol{k}_1\right)\cdot\boldsymbol{x}^{(0)}_i}
        \\
         &\qquad\times\left[1-\mathrm{e}^{-i\boldsymbol{k}_1\cdot\left(\boldsymbol{x}^{(0)}_i-\boldsymbol{x}^{(0)}_j\right)}\right]\boldsymbol{u}_{\boldsymbol{k}_1}\\
         =&-\sum_{\boldsymbol{k}_1}\sum_{\boldsymbol{x}^{(0)}_{ij}\in\mathrm{\{N.N.\}}}\underline{\boldsymbol{M}}_{ij}\cdot\delta_{\boldsymbol{k},\boldsymbol{k}_1}
         \left[1-\mathrm{e}^{-i\boldsymbol{k}_1\cdot\boldsymbol{x}^{(0)}_{ij}}\right]\boldsymbol{u}_{\boldsymbol{k}_1}\\
         =&-\sum_{\boldsymbol{x}^{(0)}_{ij}\in\mathrm{\{N.N.\}}}\underline{\boldsymbol{M}}_{ij}\cdot\left[1-\mathrm{e}^{-i\boldsymbol{k}\cdot\boldsymbol{x}^{(0)}_{ij}}\right]\boldsymbol{u}_{\boldsymbol{k}}\\
         =:&-\underline{\boldsymbol{M}}_{\boldsymbol{k}}\cdot\boldsymbol{u}_{\boldsymbol{k}}.
    \end{split}
\end{equation}
We denote the set of all lattice vectors connecting nearest-neighboring objects in the undeformed ground state by $\{\mathrm{N.N.}\}$. In the fourth equality, we switch to the sum over all $\boldsymbol{x}^{(0)}_{ij}=\boldsymbol{x}^{(0)}_i-\boldsymbol{x}^{(0)}_j\in\{\mathrm{N.N.}\}$. This is possible because both $\underline{\boldsymbol{M}}_{ij}$ and the exponent only depend on these relative distance vectors. Moreover, since $\boldsymbol{x}^{(0)}_{ij}=\boldsymbol{x}^{(0)}_i-\boldsymbol{x}^{(0)}_j$ refer to the relative distance vectors of the lattice in the undeformed state between nearest neighbors, they are identical for all lattice points, see also Fig.~\ref{fig:hexagonal}. Therefore, the sum over $\boldsymbol{x}^{(0)}_{ij}\in\{\mathrm{N.N.}\}$ can be decoupled from the sum over $i$.  
The resulting expression for $\underline{\boldsymbol{M}}_{\boldsymbol{k}}$ is the one entering Eq.~\eqref{eq:elastic_k}.

\section{Wiener--Khinchin theorem \label{appendix:wiener-khinchin}}
First, we define the function $C_{\alpha\beta}(\tau;T)$ based on the real-valued stochastic variable $x_\alpha(t)$ via
\begin{equation}
    \begin{split}
        C_{\alpha\beta}(\tau;T)=\frac{1}{2T}\int_{-\infty}^{\infty}dt \,\braket{W_T(t+\tau)x_\alpha(t+\tau)W_T(t)x_\beta(t)},
    \end{split}
\end{equation}
where $W_T(t)$ is a rectangular window function defined in Appendix~\ref{appendix:fourier}.
Moreover, we define the function $S_{\alpha\beta}(\omega;T)$ as
\begin{equation}
    \begin{split}
        S_{\alpha\beta}(\omega;T)=\frac{1}{2T}\braket{\tilde{x}_\alpha(\omega;T)\tilde{x}^\ast_\beta(\omega;T)}.
    \end{split}
\end{equation}

Here, $\tilde{x}_\alpha(\omega;T)$ is the quantity obtained from $x_\alpha(t)$ via Fourier transformation in time.
In fact, it can be demonstrated that $C_{\alpha\beta}(\tau;T)$ and $S_{\alpha\beta}(\omega;T)$ are connected to each other through Fourier transformations via
\begin{align}
    S_{\alpha\beta}(\omega;T) =& \int_{-\infty}^{\infty}d\tau \,C_{\alpha\beta}(\tau;T)\mathrm{e}^{i\omega\tau},\\
    C_{\alpha\beta}(\tau;T) =&  \frac{1}{2\pi}\int_{-\infty}^{\infty}d\omega\, S_{\alpha\beta}(\omega;T)\mathrm{e}^{-i\omega\tau}.
\end{align}

Finally, in the limit of $T\rightarrow\infty$, the function $S_{\alpha\beta}(\omega;T)$ becomes the fluctuation spectrum $S_{\alpha\beta}(\omega)$, while
$C_{\alpha\beta}(\omega;T)$ turns into the time correlation function $C_{\alpha\beta}(\tau)$. 
Therefore, these two functions satisfy
\begin{align}
    S_{\alpha\beta}(\omega) =& \int_{-\infty}^{\infty}d\tau \,C_{\alpha\beta}(\tau)\mathrm{e}^{i\omega\tau},\\
    C_{\alpha\beta}(\tau) =&  \frac{1}{2\pi}\int_{-\infty}^{\infty}d\omega\, S_{\alpha\beta}(\omega)\mathrm{e}^{-i\omega\tau}.
\end{align}
These relations are expressed by the Wiener--Khinchin theorem \cite{kubo1998statistical}.

% The \nocite command causes all entries in a bibliography to be printed out
% whether or not they are actually referenced in the text. This is appropriate
% for the sample file to show the different styles of references, but authors
% most likely will not want to use it.
% \nocite{*}
\UseRawInputEncoding
\bibliography{main}% Produces the bibliography via BibTeX.

\end{document}